\documentclass[9pt,journal,cspaper,compsoc]{IEEEtran}

\usepackage{times}

\usepackage[backref=page]{hyperref}
\hypersetup{
    colorlinks = true, 
    linkcolor = blue,
    anchorcolor = red,
    citecolor = blue, 
    filecolor = red, 
}
\hypersetup{pdfinfo={
   Author		= {Ben Kenwright},
   Title		= {Watch Your Step: Real-Time Adaptive Character Stepping},
   Subject 		= {Real-Time Adaptive Character Stepping},
   CreationDate = {D:20130530195600},
   Keywords 	= {stepping;interactive;balancing;character;animation;physics-based;responsive;adaptive;dynamic;3D;video-games;inverted-pendulum;ankle-torque;controllable},
}}

\usepackage{amsmath}

\ifCLASSOPTIONcompsoc
\else
\fi

\ifCLASSINFOpdf
\else
\fi

\usepackage{graphicx}
\graphicspath{{./images/}}

\newcommand{\figuremacroW}[4]{
	\begin{figure} 
		\centering
		\includegraphics[width=#4\columnwidth]{#1}
		\caption[#2]{\textbf{#2} - #3}
		\label{fig:#1}
	\end{figure}
}

\newcommand{\figuremacroF}[4]{
	\begin{figure*} 
		\centering
		\includegraphics[width=#4\textwidth]{#1}
		\caption[#2]{\textbf{#2} - #3}
		\label{fig:#1}
	\end{figure*}
}

\begin{document}
%

\title{``Watch Your Step''\\ \vspace{-7pt}
\fontsize{13}{28}\selectfont{Real-Time Adaptive Character Stepping}}

\author{Ben Kenwright}

\markboth{Communication Article May 2013}%
{Watch Your Step: Real-Time Adaptive Character Stepping (by Ben Kenwright)}

\IEEEcompsoctitleabstractindextext{%
\begin{abstract}
An effective 3D stepping control algorithm that is computationally fast, robust, and easy to implement is extremely important and valuable to character animation research.  In this paper, we present a novel technique for generating dynamic, interactive, and controllable biped stepping motions.  Our approach uses a low-dimensional physics-based model to create balanced humanoid avatars that can handle a wide variety of interactive situations, such as terrain height shifting and push exertions, while remaining upright and balanced.  We accomplish this by combining the popular inverted-pendulum model with an ankle-feedback torque and variable leg-length mechanism to create a \textit{controllable solution} that can adapt to unforeseen circumstances in real-time without key-framed data, any offline preprocessing, or on-line optimizations joint torque computations.  We explain and address oversimplifications and limitations with the basic IP model and the reasons for extending the model by means of additional control mechanisms.  We demonstrate a simple and fast approach for extending the IP model based on an ankle-torque and variable leg lengths approximation without hindering the extremely attractive properties (i.e., computational speed, robustness, and simplicity) that make the IP model so ideal for generating upright responsive balancing biped movements.  Finally, while our technique focuses on lower body motions, it can, nevertheless, handle both small and large push forces even during terrain height variations.  Moreover, our model effectively creates human-like motions that synthesize low-level upright stepping movements, and can be combined with additional controller techniques to produce whole body autonomous agents.

\end{abstract}

}


\teaser{%
  \includegraphics[width=1.0\linewidth]{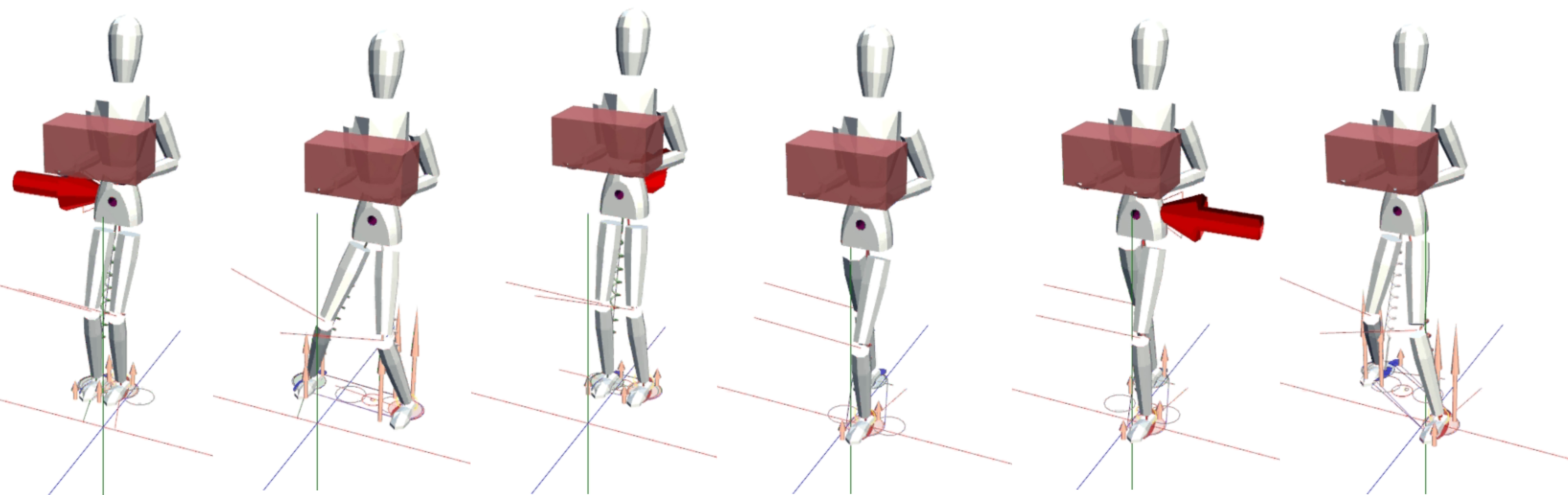}
  \caption{Our adaptive biped stepping approach ensures our character remains upright and balanced during a wide range of interactive situations.  For example, the figure shows our character holding a box, while we repeatedly apply random push forces; requiring the character to take a corrective step to regain its balance and not fall over. \vspace{-50pt}}
  \label{fig:holding_box_and_pushed}
}

\maketitle

\IEEEdisplaynotcompsoctitleabstractindextext

%


\section{Introduction}
\IEEEPARstart{C}{r}eating physically accurate, controllable, and adaptable character motions that can mimic real-world humans is challenging, interesting, and important \cite{YLP07,Ken11,BPW93,RGBC96}.  In particular, there is a great interest in developing techniques and algorithms that can run at real-time frame-rates while being robust and interactive.  The challenges stem from the fact that humans possess a huge number of degrees-of-freedom (DOF) and are capable of producing a vast assortment of diverse, original, and complex movements that are physically bound (i.e., balanced and dynamic).  While data-driven approaches using recorded human movements (i.e., motion capture data) can be played back in virtual environments to synthesize highly life-like character actions, they can be inflexible and difficult to adapt to unforeseen circumstances in real-time interactive environments, such as video games.  Then again, physics-based character models are a promising alternative to pure data-driven approaches and allow us to generate interactive motions that realistically mimic the real-world (e.g., muscles, contacts, gravity, and external forces acting on the body).  However, while the dynamics of a human can be accurately modeled using physics-based methods, \textit{implementing a real-time ``controllable'' system} that can handle disturbances, such as pushes and changes in terrain height, while synthesizing balancing motions is difficult ~\cite{KLK04,ZOHO02}.

Initially, SIMBICON \cite{YLP07} was one of the first systems to demonstrate a simple, easy-to-implement, robust physics-based controller approach for generating balanced upright stepping motions based on the biomechanically inspired, computationally-efficient, uncomplicated inverted-pendulum (IP) model.  Furthermore, due to its numerous attractive features, it later led to additional adaptations to meet other challenges, such as stepping planning \cite{SBYV08}, controller adjustments \cite{YCBV08}, and CoM-positioning \cite{SPM09}.  However, the fundamental IP model can generate a stereo-typed marching-like gait, which is primarily due to a number of oversimplifications.  For example, the low-dimensional IP model has pin-point feet contacts so it constantly needs to step to prevent itself from falling over; furthermore, when starting from a static pose the IP model needs to take multiple steps to steer a specific direction, while being unable to counter act small disturbances without stepping.


\textit{Our approach addresses oversimplifications and limitations of the basic IP model and extends the model without hindering the extremely attractive properties (i.e., computational speed, robustness, and simplicity) that make it ideal for generating upright responsive balancing biped movements.}  For example, we include an ankle-feedback torque to remedy the pin-point feet contacts.  Our modifications evolve around low-level approximation enhancements to create physically-accurate life-like biped stepping motions \textit{without} key-framed data that can be used in online environments (e.g., games).  We persistently analyze the character's parameters (e.g., center-of-mass position, comfort factors, terrain height, foot locations) to predict compensatory poses that can maintain balance during upright actions (i.e., through ankle torque, stepping movements, and leg length).  The method in this paper uses no key-framed data and requires no pre-processing; and has a minimum number of tuneable parameters.  We automatically create basic upright balancing motions, such as standing without stepping, pulling, pushing, walking up/down slopes, handling force disturbances, and dynamically changing terrain height.



\textbf{Contribution}- The key contribution of this paper is a straightforward, computationally fast, and robust physics-based method for generating `controllable' balanced upright biped stepping motions without key-frame data that are dynamic and interactive.  We accomplish this by exploiting a low-dimensional model and incorporating multiple approximation techniques (e.g., foot support area, dynamic rest leg length) to enable the character to handle disturbances in real-time, such as pushes and uneven terrain changes.  In summary, the key contributions of this paper are:

\begin{itemize}[noitemsep,topsep=0pt,parsep=0pt,partopsep=0pt]
  \item We present a practical real-time method for generating dynamic upright biped motions that are interactive (i.e., handle a variety disruptive disturbance types) and controllable (i.e., walk speed, direction, pulling, pushing objects)
  \item We incorporate an uncomplicated foot support area to emulate ankle torque, which provides a method for ``controlling'' the inverted-pendulum base model and generating a more governable and stable human avatar movement
  \item Incorporate variable leg-lengths to allow the model to handle changing terrain heights (e.g., a fixed leg-length model cannot stand still on a slope).
\end{itemize}

\section{Related Work}
There has been a broad range of exciting and interesting approaches across different disciplines (i.e., graphics, robotics, biomechanics) towards creating more interactive character animation solutions.  Whereby, we briefly review some of the most recent and relevant research in the field that has contributed to making characters motions more \emph{interactive and responsive}.


Data-driven and kinematic approaches offer a straightforward technique for creating articulated character motions in real-time.  Whereby, Komura et al. \cite{KHL05,KLK04} utilized kinematic methods by modifying movements so they emulated the responsiveness of external impacts.  Oshita and Makinouchi \cite{OMMA01} encapsulated balance and comfort factors to influence the dynamic-model in the final motions.  While Ye and Liu \cite{YYLC08} kinematically constrained and influenced humanoid character motions to respond to small perturbations.  Alternatively, hybrid solutions have switched been kinematic and physics-based approaches \cite{ZMCF05,SPF03}.  Feng et al. \cite{FXS12} demonstrated motion synthesis of object manipulation and locomotion.  However, our method is built on top of a low-dimensional physics-based model for generating the fundamental dynamic movements and uses kinematic methods for reconstructing the final articulated character pose.  

Physical controller models are a common method for creating responsive motions.  For example, different controller models have been developed for specific body regions, such as upper body movements by Zordan and Hodgins \cite{ZOHO02} and later by Yin et al. \cite{YCP03}, standing \cite{ASP07}, moving and controlling \cite{ABPJ06}, cyclic motions (e.g., walking) \cite{Ken11,SKL07,YLP07}.  A popular approach by researchers for creating physics-based animations is using simplified machines to generate base motions (e.g., the inverted-pendulum (IP) for locomotion) \cite{PCTD+01,KKKF+03,TLCL+10,CBM10}.  Additionally, Kenwright \cite{Ken12} demonstrated a real-time technique for producing responsive balancing motions by extending the IP model with enhancements, which we adopt in our approach to create our reactive character solution.  However, our method focuses on extending the low-dimensional physics-based model to accomplish a more flexible solution that can also \textbf{adapt do terrain height changes} and force perturbations (e.g., pushes) in real-time.  Energy efficient factors for walking motion generation and adaption \cite{WFH10,WHDK12}.  Coros et al. \cite{SBYV08} and Wu and Zordan \cite{WZ10} present physics-based controllers that are aware of their foot plan and try to follow it as closely as possible, while Mitake et al. \cite{MAAM09} generates character motions using a multi-dimensional key-frame animation in parallel with a physical simulation.  We show a visual comparison of overlapping work that focused on similar goals based on parallel methods (e.g., IP model) in Figure \ref{fig:comparison_grid}.


\figuremacroF
{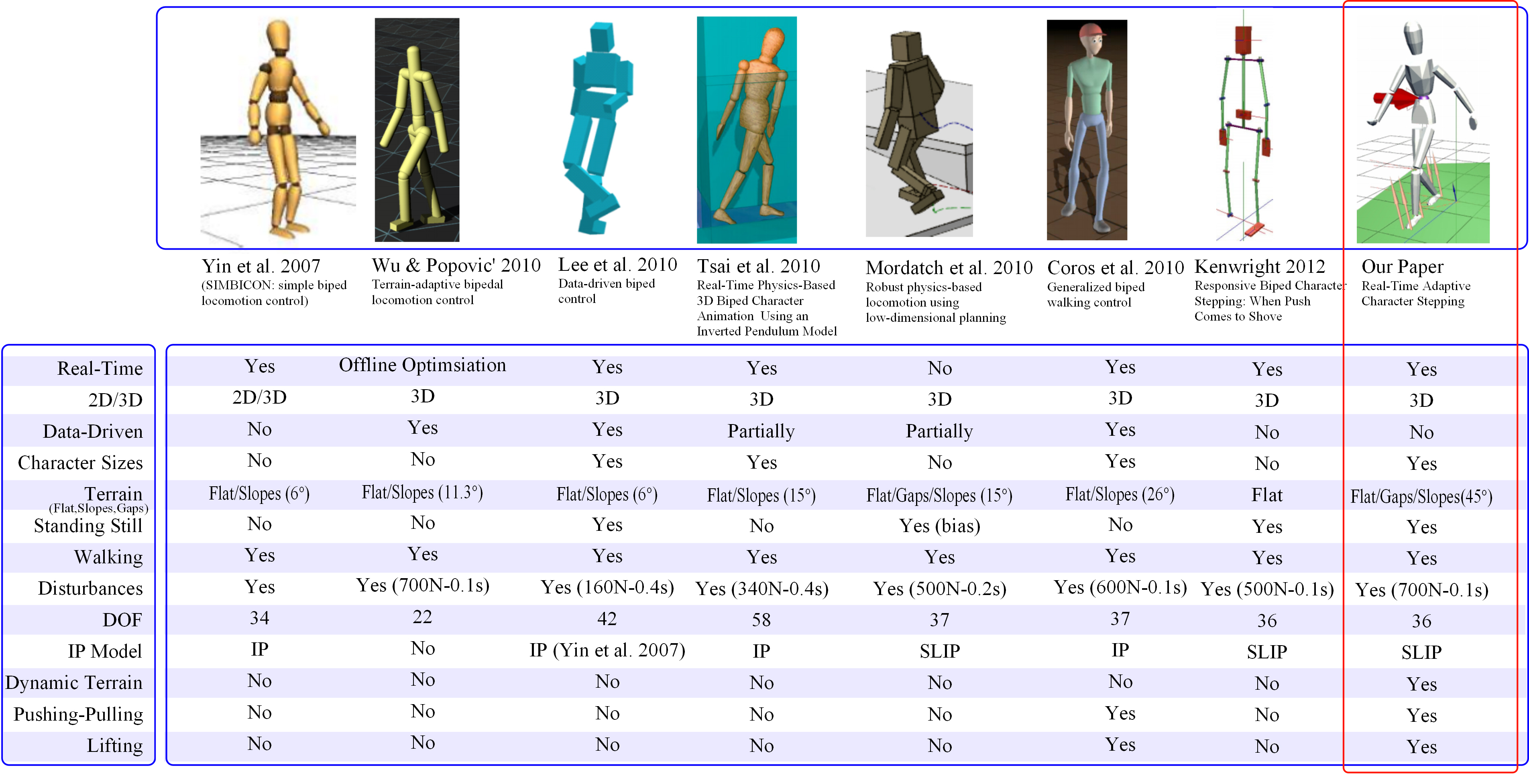}
{Comparison Grid}
{We show a comparison grid to emphasis our approaches similarities and differences with existing publications that have approached the problem of upright balanced biped character motions with similar techniques (e.g., IP model) enabling the reader to see at a glance the crucial differences.}
{1.0}


Emphasizing some of the relevant work in the field of robotics that contributed to the development of responsive biped controllers, we outline a few interesting and important papers.  Firstly, there was the impressive work by Shih et al. \cite{SGL93} who developed a straightforward model for enabling biped character walking machines to respond to small disturbances.  Secondly, there was the work by Pratt et al. \cite{PCD06} and Stephens \cite{Step07} who developed controllers, which could generate motions for recovering from a range of push disturbances.

Finally, we should credit the significant work by NaturalMotion, who provide a proprietary middleware physics-based character engine called Eurphora \cite{NATU12} that successfully created balancing bipeds in real-time for the video game industry (i.e. it was used in the game `Grand Theft Auto IV' \cite{ROCK12}).

\figuremacroW
{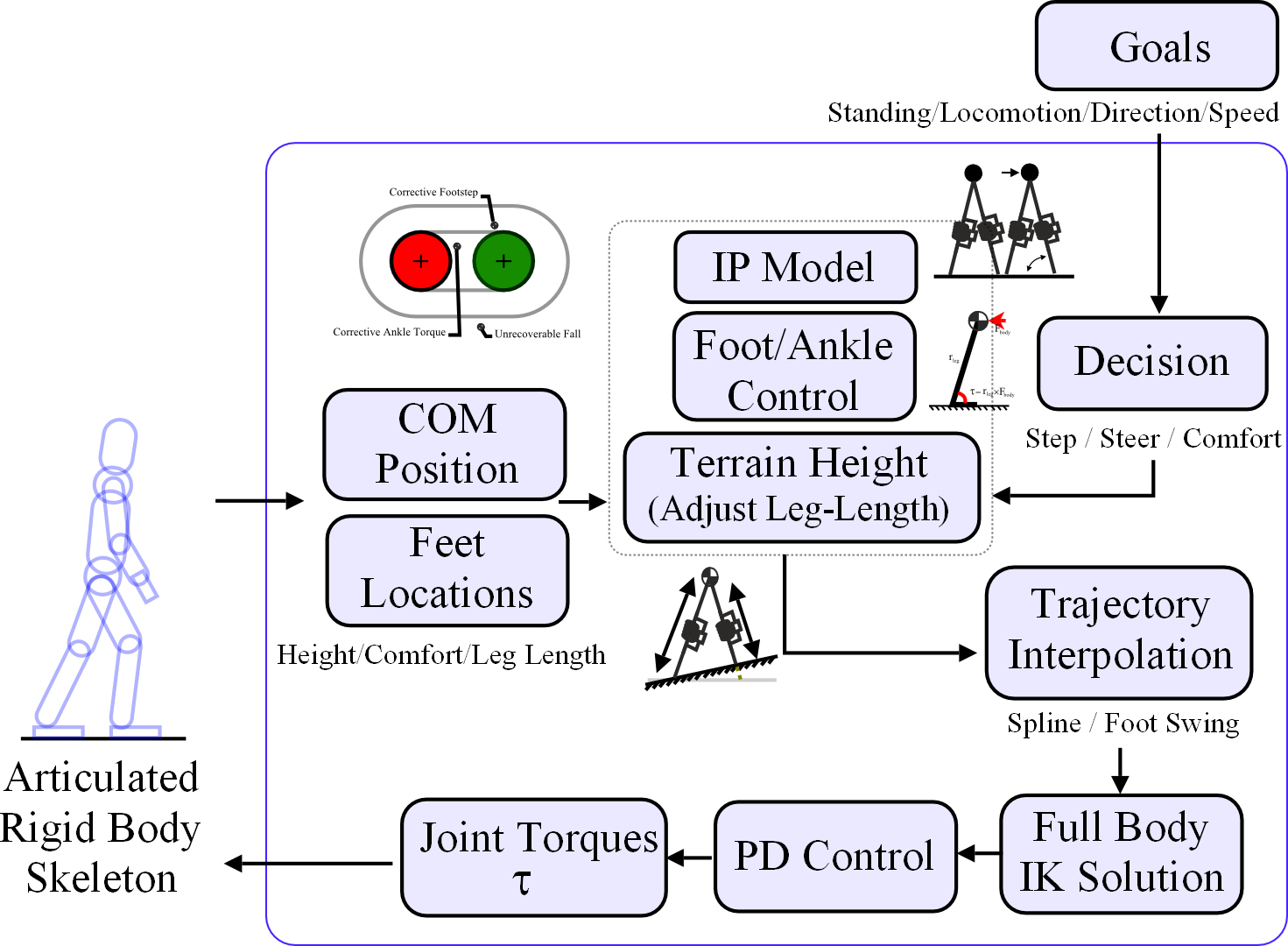}
{System Overview}
{Iteratively the character's state (i.e., feet and COM position) are analyzed to determine the next course of action.  The desired goal, balancing situation, and comfort factors all contribute to the final pose which is used to calculate the rigid body joint torques for the resulting articulated skeleton.}
{1.0}

\section{Overview}
Our algorithm uses a trouble-free state-machine logic to iteratively generate and correct fundamental upright motions (e.g., standing and walking).  Whereby, our model analyzes the current balancing situation based on the center-of-mass and foot placement information thereupon making an intelligent decision on how to proceed (e.g., add corrective ankle torque, re-position feet).  However, to accomplish this efficiently and in real-time frame-rates, our model exploits multiple approximations (e.g., circular foot support area, low-dimensional physics-based model) so we are able to synthesize upright biped character motions.  Moreover, due to the dynamic nature of our model, it can create motions for different disturbances (such as pushes, terrain height changes) which are aesthetically pleasing (e.g., the motions are highly responsive and mimic human-like stepping movements).  An overview of our system is outlined in Figure \ref{fig:overview}.

Multiple interesting and inspiring research has contributed to our model and demonstrate that each of our model's components (proportional derivative (PD) controller, inverted-pendulum (IP), simplified foot support area, variable leg lengths) have at one point all been successfully used in the field of animation.  Whereby, each of the different components has been exploited in one form or another over the past decade.  We, nonetheless, combine them to focus specifically on creating controlled responsive and adaptive biped stepping motions.  We have, however, to the best of our knowledge never seen them combined in this approach for generating responsive motions that can handle dynamically changing terrain height and force perturbations (i.e. pushes and hits).
\section{Fundamentals}

\figuremacroW
{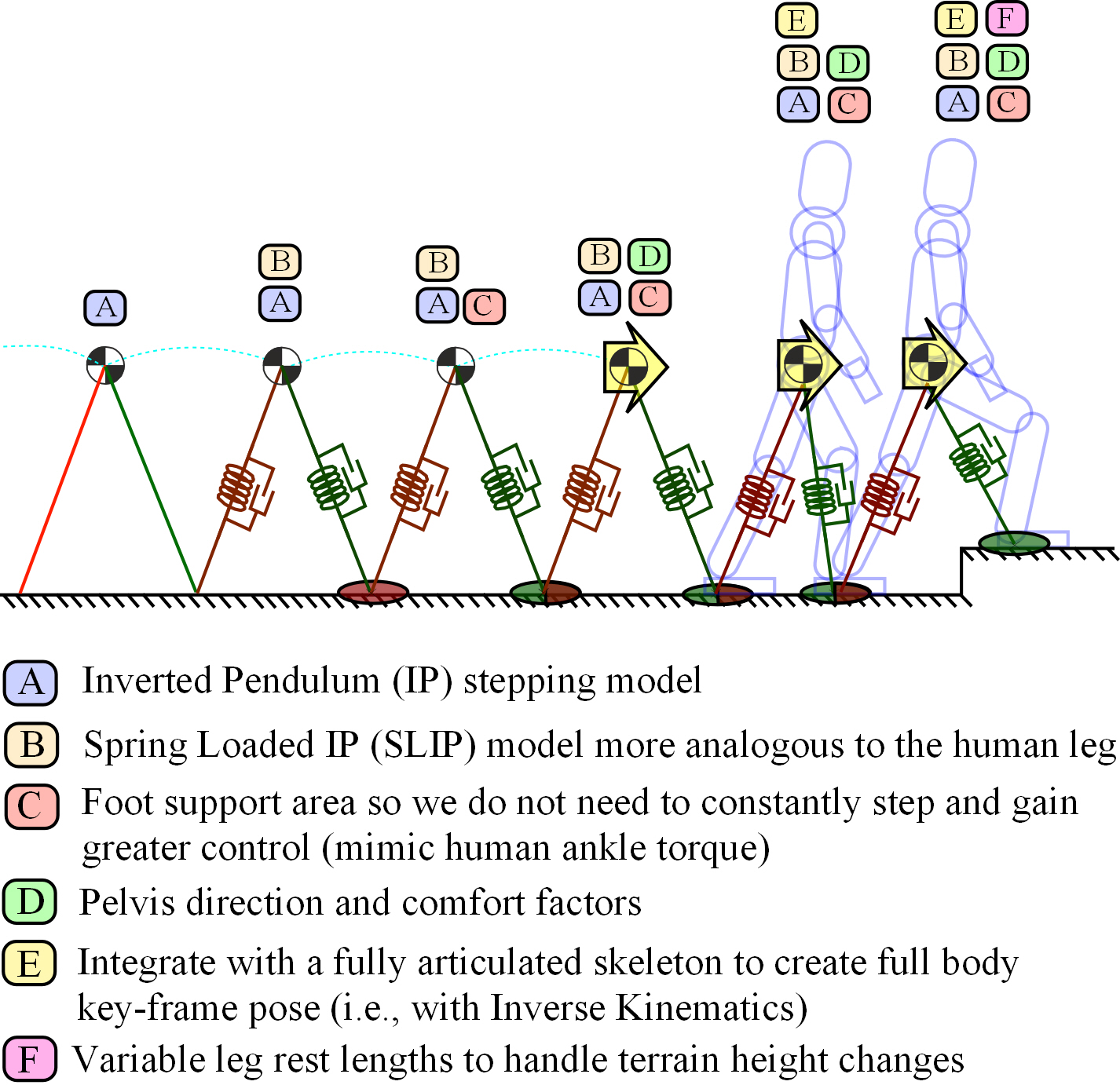}
{Evolution}
{The evolutionary reasoning behind our model, which starts from the popular inverted-pendulum (IP) model that provides a highly robust, efficient, and straightforward approach for generating balancing human foot placement information towards our adapted more human-like model that exploits all the advantages of the IP model but provides multiple feature enhancements (e.g., simplified foot support area for ankle-torque, leg length deviations, and steering).}
{0.9}

\subsection{Inverted-Pendulum}
The Inverted-Pendulum (IP) ~\cite{KKKY+01} is a popular biomechanically inspired technique for predicting biped character stepping motions ~\cite{KLK04,TLCL+10}.  Whereby, the IP is a low-dimensional model that represents the total character's mass as a point, which is supported upon a massless leg.  Nevertheless, this simplified model is able to capture significant dynamical aspects of a biped's motions ~\cite{CBM10}.  However, the leg is more akin to a spring-damper mechanism; hence, we adopt the Spring-Loaded Inverted-Pendulum (SLIP) ~\cite{FUKO99,Ken11} model (shown in Figure ~\ref{fig:evolution}) as the building block for generating our stepping information.

The IP is an intuitive, abstract, and flexible approach for identifying and generating balancing information.  Furthermore, it can be used for different motion styles and simplifies the multi-segment skeleton down so that we do not need to perform complex kinematic and dynamic calculations.  The fundamental IP model works by pole vaulting the point-mass over the support leg during locomotion while transitioning instantly between the left and right leg between stepping.  Hence, energy is constantly converted between kinetic and potential energy to generate constant perpetual stepping motions through the conservation laws of mechanical energy (i.e., as shown by the fundamental Equation ~\ref{eq:potential_kinetic}, whereby the sum of the potential and kinetic energy should be conserved before and after the stepping phase).  This principle can be used to estimate the character's foot placement position (e.g., step length) to control the speed (e.g., walk faster or come to a stop) and direction while maintaining balance, see Coros et al. ~\cite{CBM10} and Tsai et al. ~\cite{TLCL+10}.

\begin{equation} \label{eq:potential_kinetic}
\frac{1}{2}mgh + \frac{1}{2}m{v^2} = \frac{1}{2}mgh' + \frac{1}{2}m{v^{'2}}
\end{equation}

\noindent where $m$ is the mass, ${g}$ is the gravitational constant (i.e., 9.8$m/s^2$), ${h}$ and ${h'}$ are the heights before and after the stepping phase, ${v}$ and ${v'}$ are the velocities before and after the stepping phase.

From which we develop the Equation \ref{eq:calculate_d_stairs} with reference to Figure \ref{fig:step_stairs}.
\begin{equation} \label{eq:calculate_d_stairs}
\begin{alignedat}{2}
	v' &= 0\\
	h' &= L = \sqrt{ ( (h-s)^2 + d^2 ) }    &&  \; \text{(Sub in Eq. \ref{eq:potential_kinetic})} \\
	\frac{1}{2} m v^2 + m g (h-s) &= m g (\sqrt{ ( (h-s)^2 + d^2 ) }) && \; \text{(Solve for d)}\\
	d &= \frac{v}{2g}\sqrt{(v^2 + 4g(h-s)}
\end{alignedat}
\end{equation}


\figuremacroW
{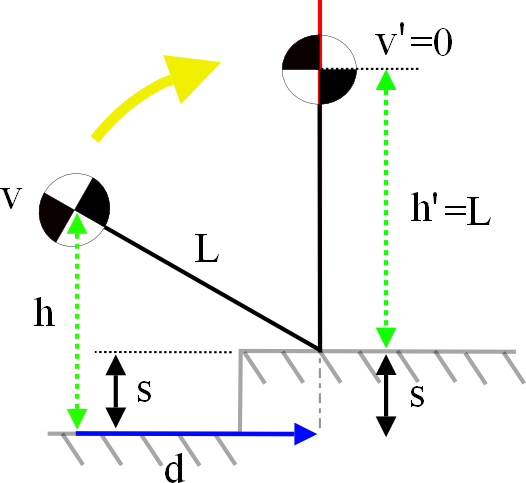}
{Sloping Terrain}
{Extending the basic model to include deviation for changes in terrain height (e.g., uneven terrain, such as stairs or sloping ground).}
{0.5}

We define the condition for the IP model to either halt or move in the desired direction (i.e., ${(\hat{u} \cdot  \hat{a}) || F_g || \ge (\hat{b} \cdot F_p )}$, as shown in Figure \ref{fig:ip_pull}).  So that the model can pull or push objects, we add a bias factor $\beta$ to the equation.  This adds to the stepping distance, as shown in Equation \ref{eq:pull_force} to include pull (or push) force in the stepping distance to maintain a controllable upright velocity in the desired direction (as shown in Figure \ref{fig:sim_results_basic_ip}(f)-(g)).

\begin{equation} \label{eq:pull_force}
\begin{alignedat}{2}
	\beta &\ge \frac{(\hat{b} \cdot F_p )}{(\hat{u} \cdot  \hat{a})|| F_g ||}\\ 
	d' &= d + \beta k_p\ 
\end{alignedat}
\end{equation}

\noindent where $k_p$ is a scaling factor (e.g., 0.4), $d$ is the stepping distance calculated using Equation \ref{eq:calculate_d_stairs}, $F_p$ is the external pull or push force exerted on the body, and $\hat{u}, \hat{a}, \hat{b}$ are unit vectors as shown in Figure \ref{fig:ip_pull}.  Hence, we modify the stepping distance Equation \ref{eq:calculate_d_stairs} so that it can continue to remain balance and walk while under the influence of push or pull forces (e.g., while dragging a box) as shown below in Equation \ref{eq:pull_force}.  Whereby, the pull (or push) force reduces and increases the step distance sufficiently to compensate for the pull (or push) and maintain relatively constant walk speed in the desired direction.  Figure \ref{fig:sim_results_basic_ip} shows velocity plots for Equation \ref{eq:pull_force} demonstrating the relatively robust and controlled nature of the formula.

\textit{It is important that the fundamental IP model remains stable on its own}.  For example, the model should be able to maintain a `constant' controllable walk speed that is robust, while being able to compensate for disturbances, as shown in Figure \ref{fig:sim_results_basic_ip}.  The feedback control mechanics (i.e., foot-ankle torque and variable leg-length) are built on top of the IP model's foundation; hence, it should be solid and dependable.  When reproducing the results, it is crucial that the basic IP model's (i.e., model without feedback or adjustments) coefficients are done first (i.e., the ankle-torque and leg-length are not a means of stabilizing the IP model only for introducing additional control and remedying over-simplifications, discussed in Section \ref{sec:inv_pend_limitations}).

\figuremacroW
{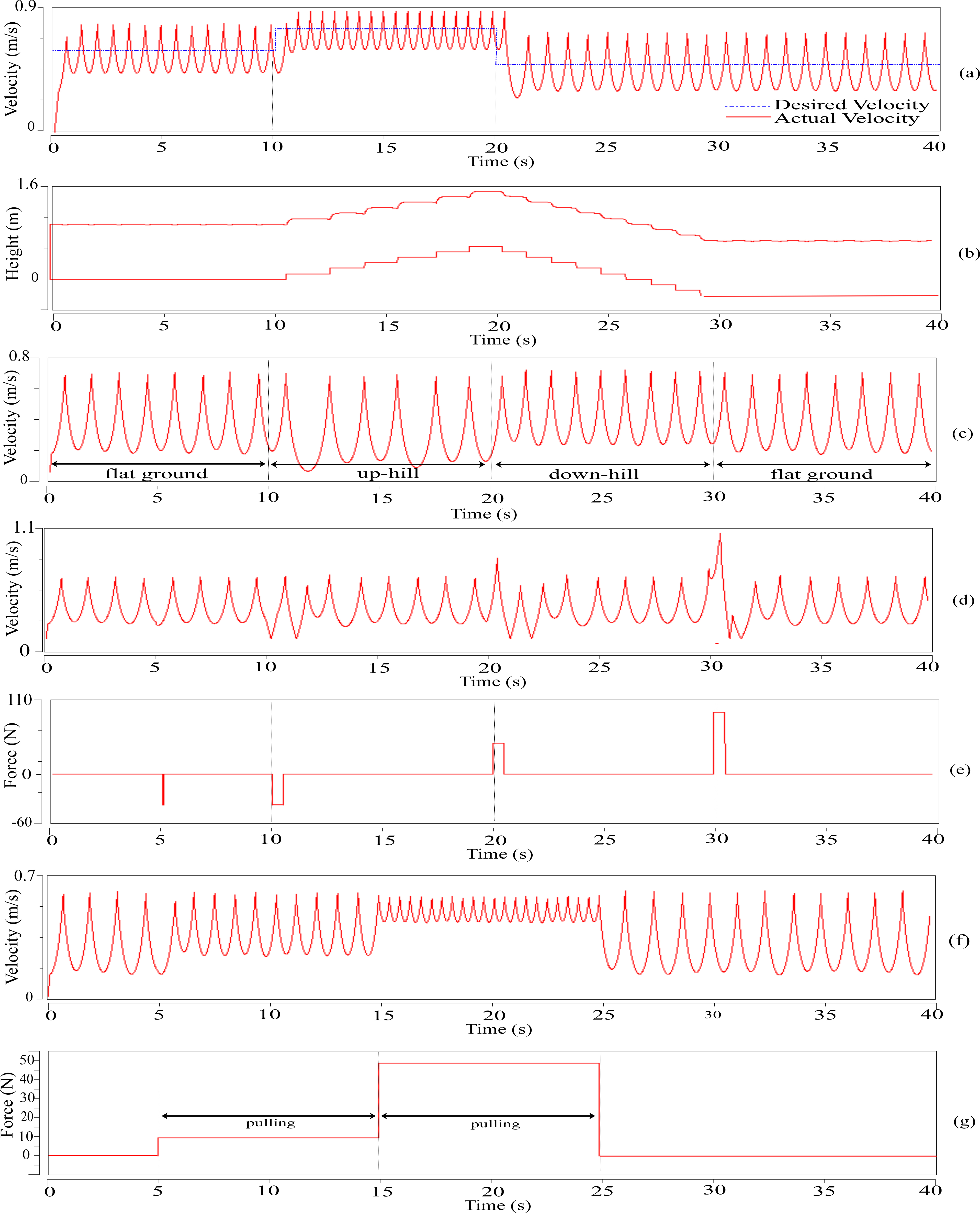}
{Simulation Results}
{Controlled walk speed under different situations. (a) walking at a constant velocity (e.g., speeding up and slowing down as we desire) by means of corrective step distance, (b)-(c) walking up-down slope while maintaining controlled speed, (d)-(e) impulse force applied but compensating and controlled walking speed, (f)-(g) pushing and pulling force while maintaining a constant speed by adjusting step distances.}
{1.05}


\figuremacroW
{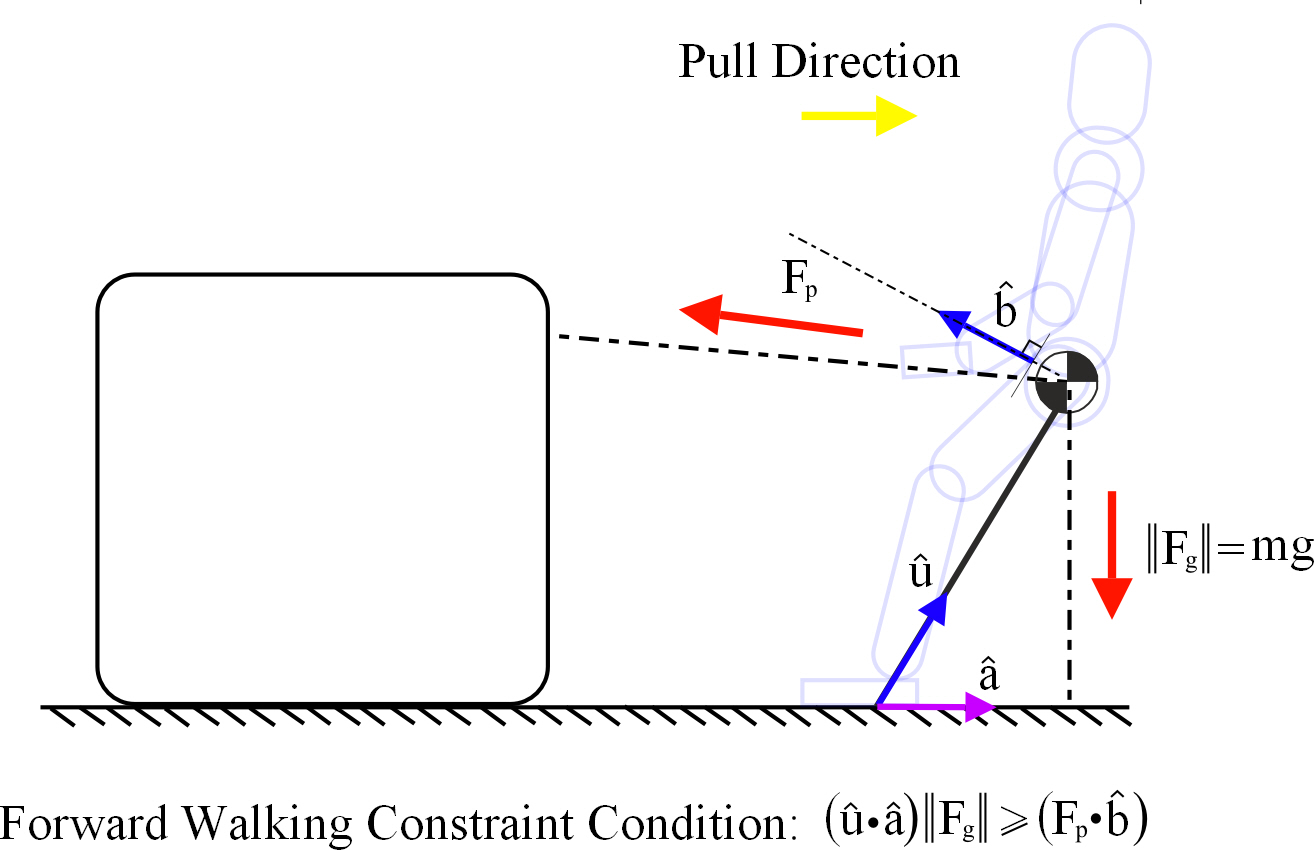}
{Pulling \& Pushing}
{We incorporate the ability to pull objects by incorporating a bias feedback control.}
{0.7}


\figuremacroW
{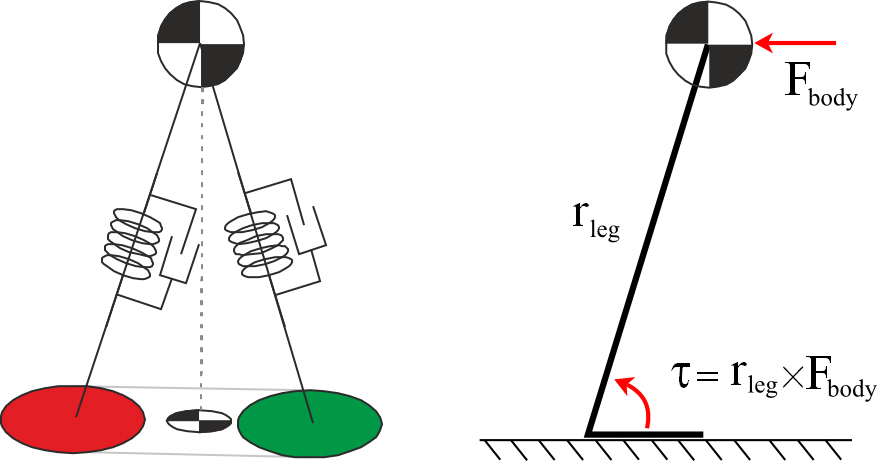}
{Model Variables}
{(Left) The foot support area using circles projected onto the ground plane, and (Right) foot support area allows us to inject ankle torque into the basic model to control the position of the center-of-mass.}
{0.6}

Since the SLIP presents an ideal method for emulating a character's leg, due to the human muscle being mechanically analogous to a spring-damper system; whereby, stiffness and damping factors can be calculated to mimic a person's limb and how they would respond.  Where Equation \ref{eq:hook} shows the spring-damper formula for representing the character's leg compression and stiffness.

\begin{equation} \label{eq:hook}
 F =  - {k_s}x - {k_d}v\
\end{equation}

\noindent where k$_{s}$ and k$_{d}$ are the spring and damping coefficients, x is the displacement vector from its rest length, and v is the relative velocity between the two points.

Equation \ref{eq:calculate_d_stairs} is based on a rigid leg and must be modified to account for the energy stored in the SLIP model.  The SLIP model contains damping that reduces oscillation and incorporates loss into the system.  The leg spring length error is used as a bias factor for $d$ so the motion remains perpetual and stable. 
\begin{figure}[htbp]
  \centering
  \includegraphics[width=1.0\columnwidth]{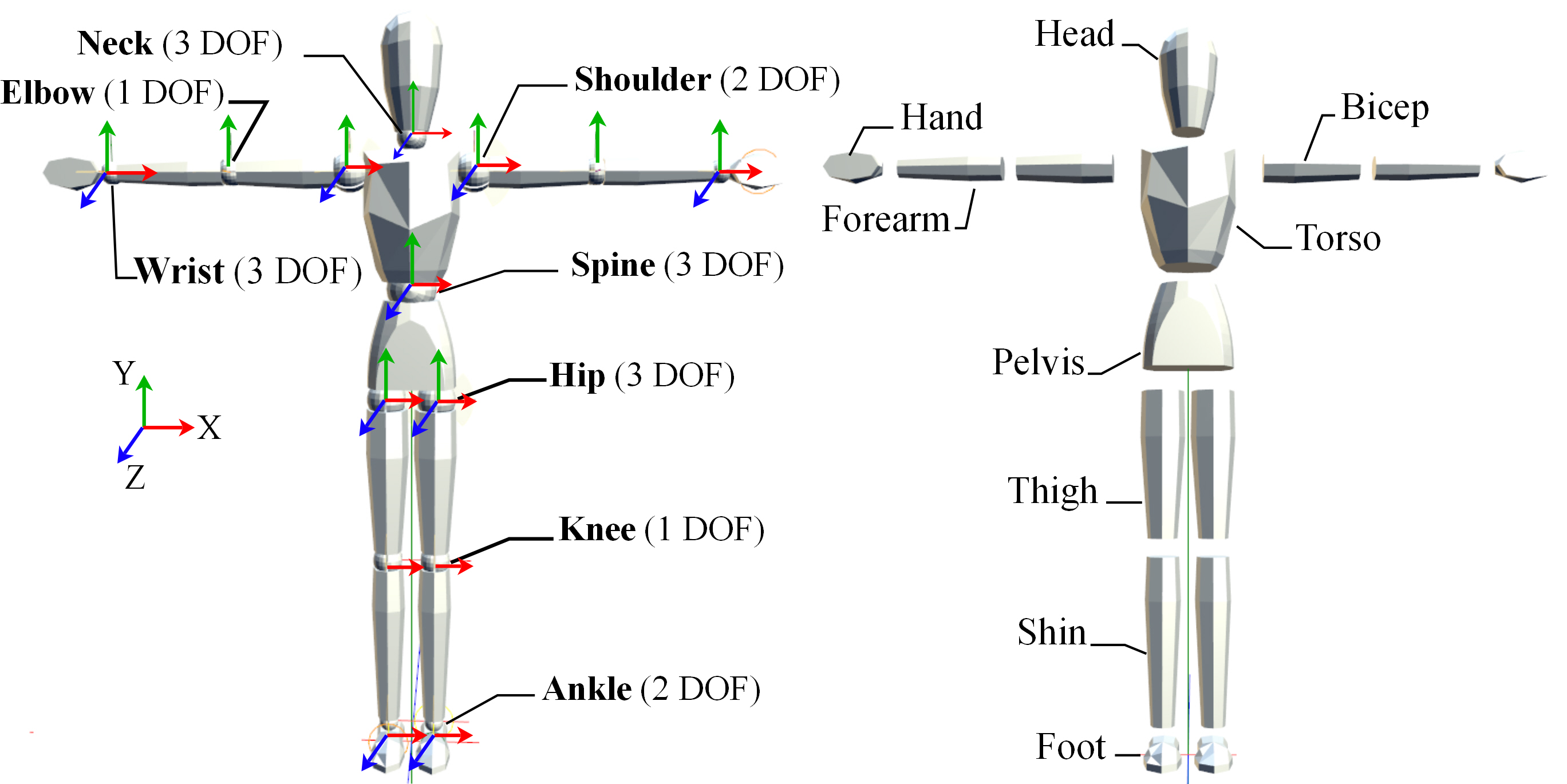}
  \caption{Character joint names, individual body parts names, and degrees-of-freedom (DOF).}
  \label{fig:joint_names}
\end{figure}

\begin{table*}
  \centering
  \includegraphics[width=0.7\textwidth]{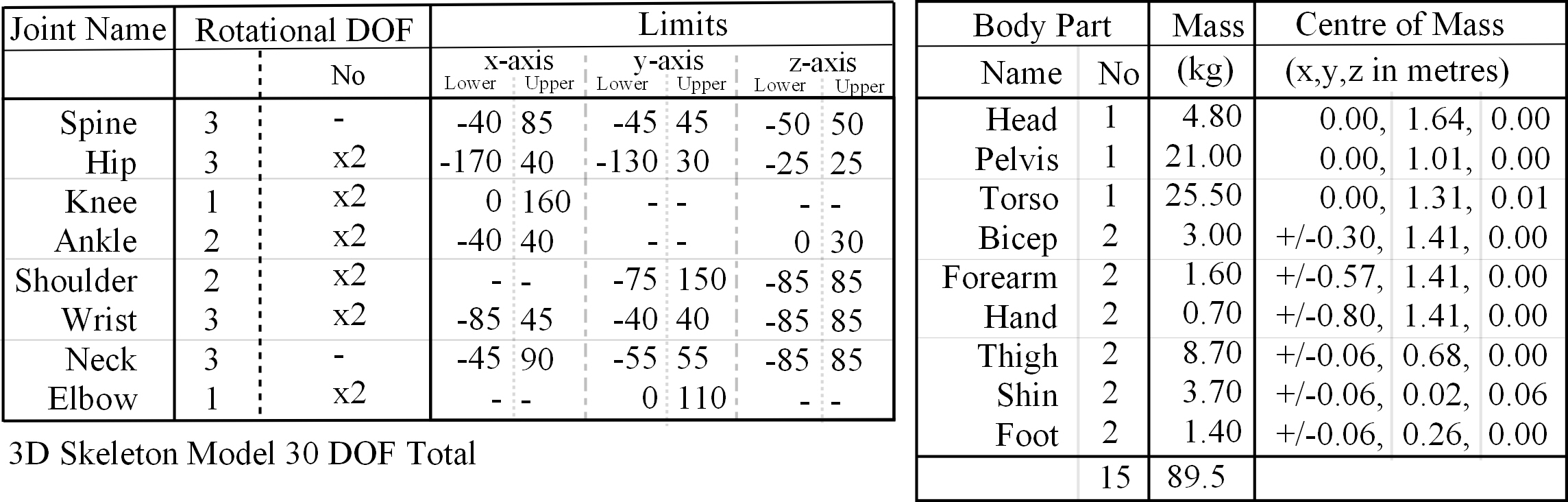}
  \caption{A list of the local 3D skeletons 30 degrees-of-freedom (DOF) limits (note an additional 6 DOF from the world root, and angular limits are shown in degrees).}
  \label{table:joint_limits}
\end{table*}


\subsection{Inverted-Pendulum Limitations} \label{sec:inv_pend_limitations}
While the IP model is computationally fast, straightforward, and responsive, it is not infallible and has a number of oversimplifications.  These fundamental oversimplifications are:
\begin{itemize}[noitemsep,topsep=0pt,parsep=0pt,partopsep=0pt]
	\item Continual state of motion (i.e., always needs to keep stepping to remain upright and balanced)
	\item Pin-Point Feet (i.e., no support area or angle torque)
	\item No feet or pelvis orientation information
	\item No postural information (i.e., upper body orientation)
	\item Mass-less legs
	\item No feet trajectory information (e.g., height, speed, orientation)
	\item Requires multiple steps for steering (i.e., cannot start locomotion from a stop and needs to wait for gravity to pull it forwards which can be the wrong desired direction) - no steering control
	\item Does not account for double support foot placement (i.e., when both feet are on the ground supporting the body)
	\item No friction or ground-feet slipping
\end{itemize}
However, we can resolve a number of the IP model's problems by including an additional control mechanism (i.e., a foot-ankle feedback torque) as we discuss later.

\subsection{Character Model}
Our character model, shown in Figure ~\ref{fig:joint_names}, is 1.7m tall and weights 89.5kg.  The kinematic and dynamic properties of the character model are shown in Table ~\ref{table:joint_limits}.  Our character possesses a total of 36 DOF (see Figure ~\ref{fig:joint_names} for details) to represent our generated motions.  We use simplified bounding boxes for collision detection instead of the model's triangle primitives.

If constant spring-damper coefficients are used for calculating the joint torques, it can produce stiff-looking motions ~\cite{LYPS*10}.  Hence, to mimic the human muscle more accurately and produce more natural-looking motions, we adopt a similar approach to Liu et al. \cite{LYPS*10} and Zordan et al. \cite{ZMMS+07} and scale the spring-damper strengths for each joint by their associated joint inertia-orientation (i.e., an inertias-scaled PD servo).

\section{Implementation}
We use different techniques to create a character that can handle a diverse set of disturbances in real-time.  We use the Open Dynamics Engine (ODE) version 0.11 ~\cite{ODE12} for simulating our character's dynamics (i.e., rigid bodies, contacts, joints) with a time-step of 0.01s, gravitational constant of 9.81$ms^-1$, and frictional coefficient ${\mu}$=1.0 (see ~\cite{ODE12} for details).  All the simulations run in real-time and were written as a single-threaded application.  Furthermore, we did not perform inter-body collisions belonging to the same character.

\figuremacroW
{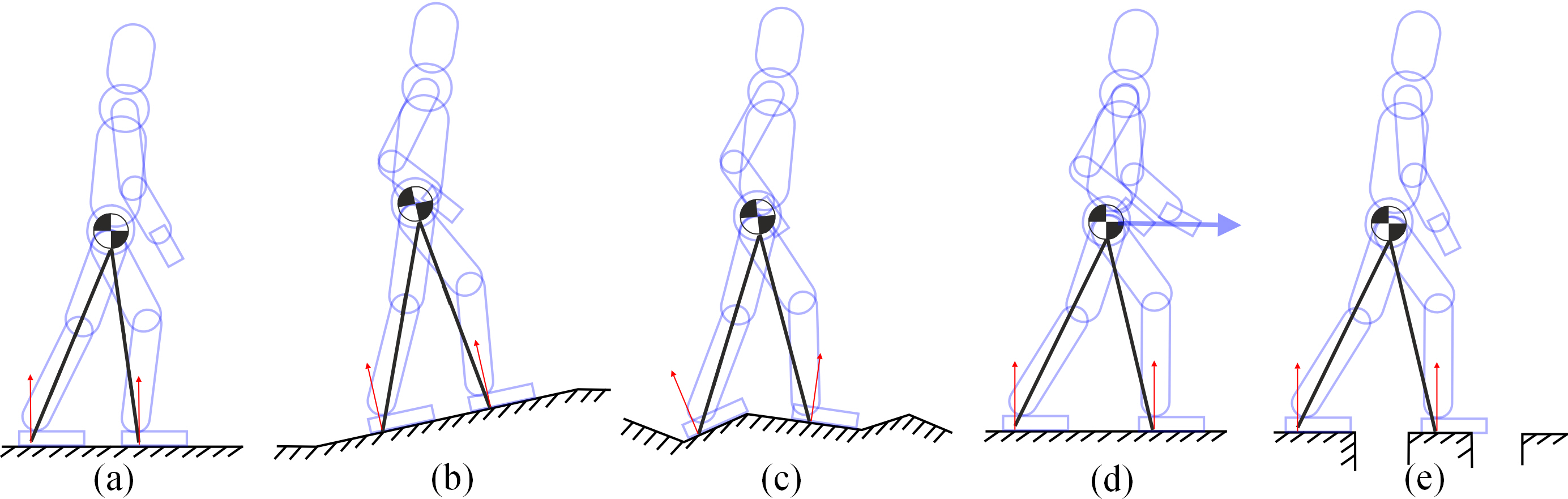}
{Stepping Model Capabilities}
{Walking at a controllable speed under a variety of conditions, such as pulling/pushing forces and terrains height deviations that includes inclines (approx. 25 degrees) and gaps (approx. 0.2m).  (a) Flat terrain, (b), sloping terrain (e.g., stairs and hills), (c), uneven ground, (d) constant pulling or pushing force, and (e) terrain gaps.}
{1.0}

\subsection{Support Area (Feet)}
The IP is a robust, computationally fast, and straightforward technique for generating responsive, dynamic, realistic looking stepping solutions.  Nevertheless, due to the model's oversimplifications, it has a number of drawbacks.  Primarily, it cannot stand still (i.e., \textit{without feet the stilt-like character must remain in a continual state of motion to maintain balance}) (see Sias and Zheng ~\cite{SIZH90} for further details on why we need feed).  Nevertheless, without compromising the advantages of the IP model, we add a support region approximation that mimics ankle torque to remedy these problems and incorporate additional control features.  We project a simplified foot support region on to the ground using spheres and capsules to keep the model computationally fast and straightforward (see Figure ~\ref{fig:ground_region}).  This simplified support region allows us to add a feedback force to the overall center-of-mass (COM) to provide control.  Hence, the ankle torque can remedy small disturbances without needing to take corrective steps.  Furthermore, the ankle torque can be used to steer the character during locomotion.  Finally, the ankle torque enables us to maintain balance on a single foot (i.e., induce feedback force on the COM to shift it between the left and right support foot as desired).


\figuremacroW
{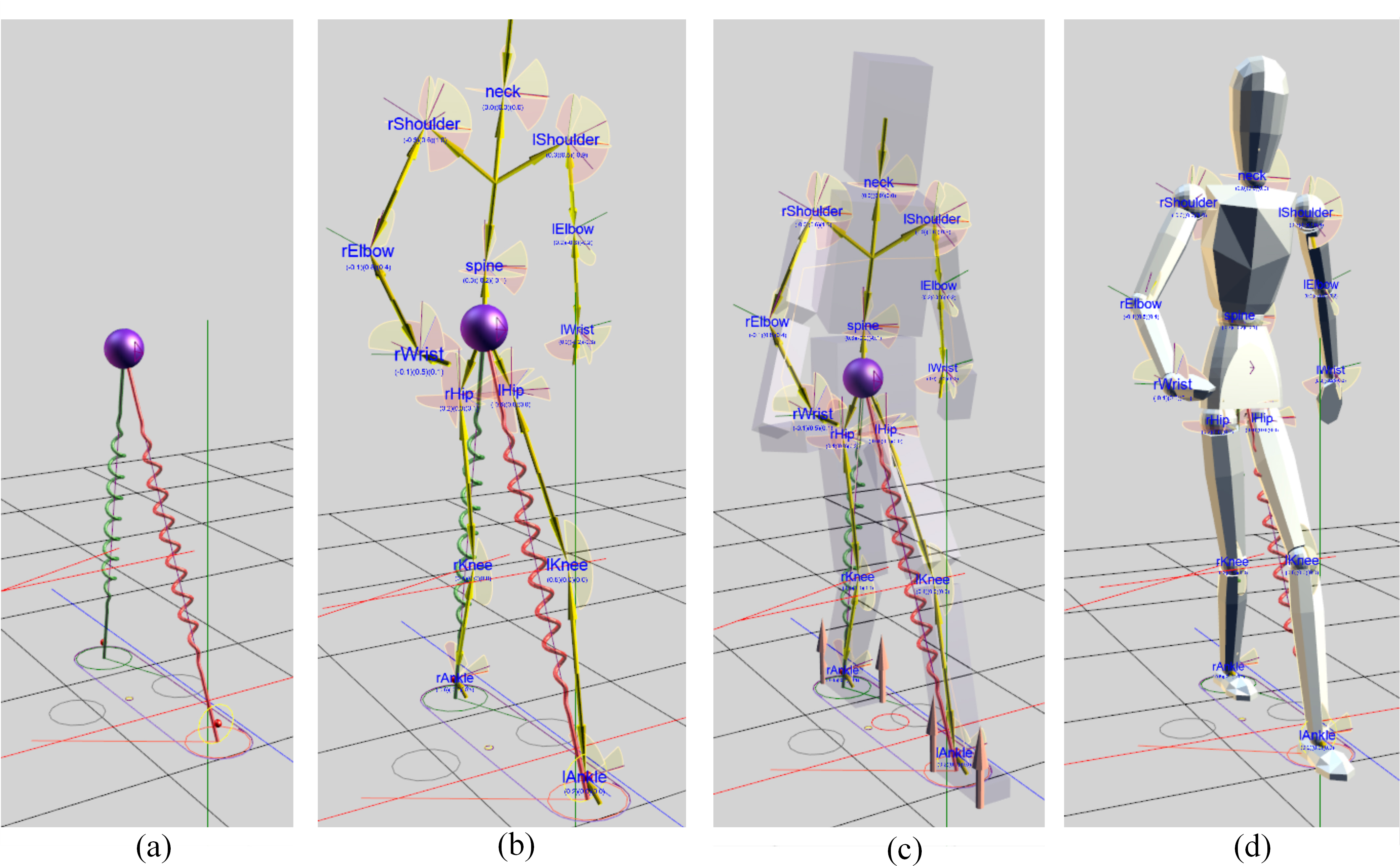}
{Decomposition}
{A visual decomposition of the character model into its separate components; (a) Stepping Model, (b) Inverse Kinematic Solution (with Joint Limits), (c) Interconnected Rigid Bodies (with Contacts), (d) Graphical Solution.}
{1.0}

\textit{It is vitally important to understand that the inverted-pendulum on its own does not provide a viable biped animation solution.  The model has a number of limitations and needs to be combined with an additional controller mechanism (e.g., elongated body for momentum feedback, pressure-point shifting, feet/ankle-torque feedback). to make it a usable tool for generating animations}

While the foot of the inverted-pendulum is typically a fixed pin-point location on the ground ~\cite{KMS01,FUKO99}, Mordatch et al. ~\cite{MLH10} takes a novel approach of moving this contact point around within the foot support region to mimic the center-of-pressure shifting and gain greater control and balance (i.e., known as foot rolling ~\cite{ACK06} ).  However, our method keeps the inverted-pendulum model foot location fixed and induce a control force from the simplified support region to mimic ankle torque to induce a controlling force on the center-of-mass (e.g., see Figure \ref{fig:model_basics}).

\subsection{Leg Lengths}
The fixed IP model works well on flat terrain but is unable to adapt to sloping or uneven ground, such as hills and stairs.  Since the SLIP model is based on the spring-damper Equation \ref{eq:hook}, we can dynamically modify the legs' rest-lengths between step transitions to accommodate terrain height changes.  Furthermore, since the leg's rest-lengths are constantly being updated based on the ground situation it means we are able to have the character adapt automatically in real-time to dynamically changing ground terrain.  For example, a slowly rotating platform shown in Figure \ref{fig:teaser};
whereby the character mimics shifting balance between the left and right leg as the height of the ground gradually changes.  We accomplished this by modifying the desired rest-length of the legs, as shown in Figure \ref{fig:resize_leg}.  Each leg's desired rest-length are calculated based on the contact positions of the feet and their heights on the terrain with respect to each other.  The longest leg is always set back to the original rest-length so that we can accomplish tasks, such as standing still on sloping terrain, in addition to climbing stairs, as shown in Figure \ref{fig:legs}.  Moreover, while the spring stiffness (and damping) are typically fixed, the approach of dynamically altering the legs' rest-lengths was also done by Mordatch et al. ~\cite{MLH10} to gain additional control benefits (e.g., ability to handle terrain height changes), so parallel and perpendicular movements to the ground plane could be achieved.


\figuremacroW
{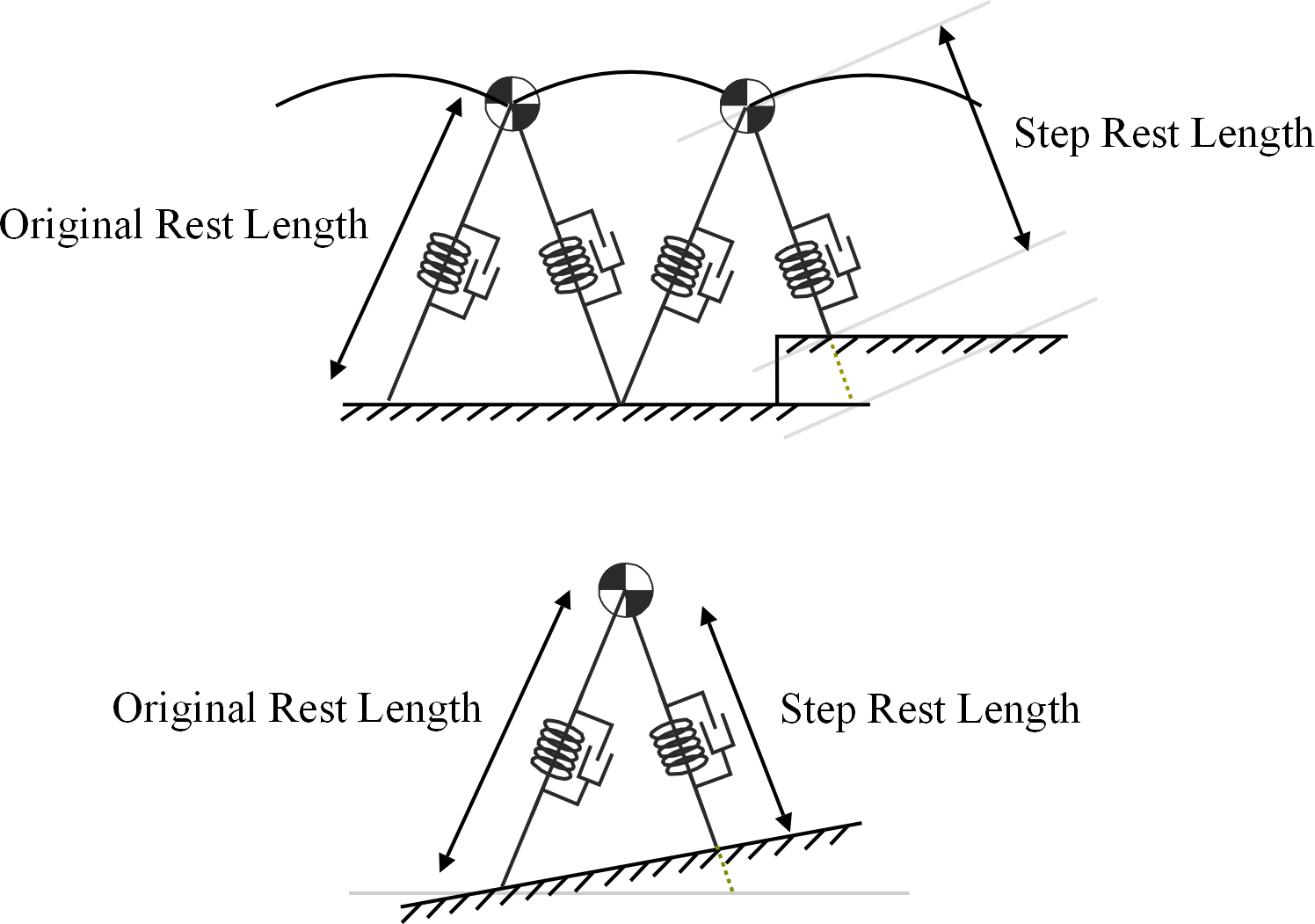}
{Uneven Terrain}
{Our model adapts to handle dynamically changing terrain heights by constantly adjusting the leg's rest-length during step transitions; returning the leg to the original rest-length to mimic the muscle extending and pushing off the weight onto the alternate leg.}
{0.82}


\figuremacroW
{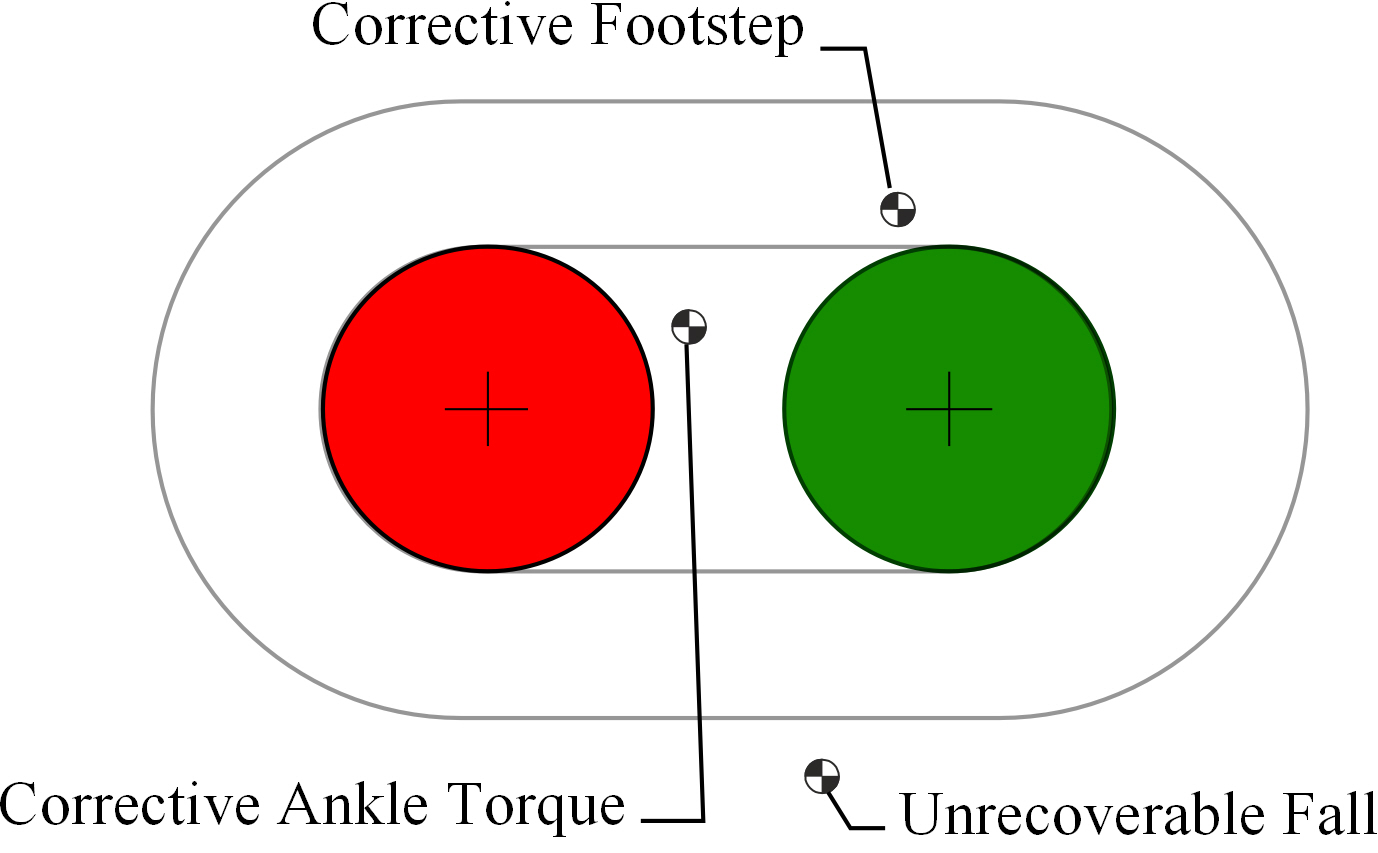}
{Stepping Logic}
{We create three regions onto which we project the center-of-mass in order to make the decision on what response is needed.}
{0.7}


\figuremacroW
{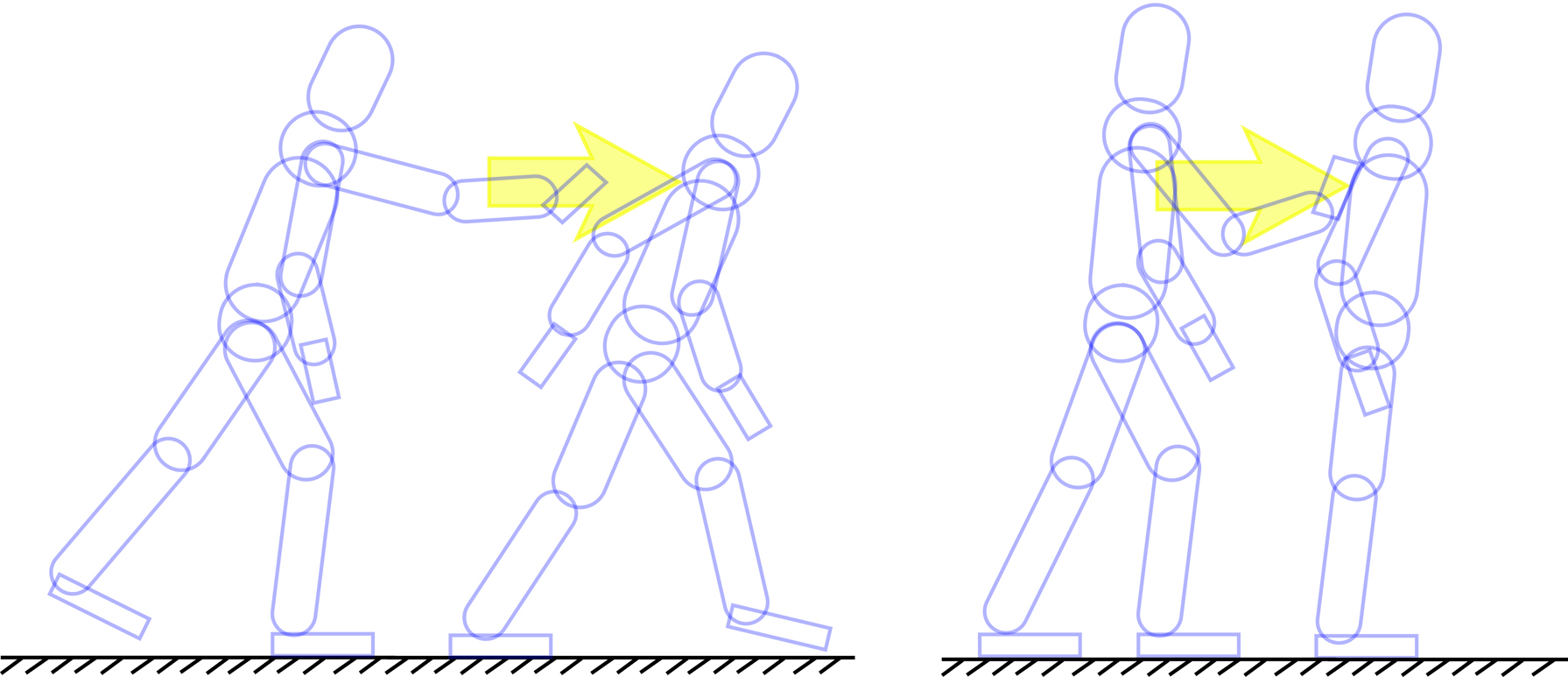}
{Disturbances}
{Illustrating a push force from a character's hand; where the magnitude causes either the character to sway (Right) or for a large enough force the character needs to take a corrective balancing step (Left).}
{0.9}

\subsection{Joint Torques}
For the tracking controller mechanism we use the popular proportional-derivative (PD) method shown in Equation \ref{eq:pd} to calculate the joint torques by minimizing the displacement between the current and reference joint angles.  Whereby, this approach has been successfully used in the past to create walking motions ~\cite{PANN96,LPF96}, athletic motions ~\cite{HWBB95}, and reactive motions ~\cite{OMMA01,ZOHO02}.  The PD controller uses the generated skeleton pose and elastic-damper coefficients to determine the necessary virtual torques to apply to the articulated rigid body biped that emulates muscles to create the final movements.  Finally, the joint torque for each of the character's joints was limited to a maximum of 500Nm. 

\begin{equation} \label{eq:pd}
 \tau  = {k_p}(\theta  - {\theta _d}) + {k_d}(\dot \theta  - {\dot \theta _d})
\end{equation}

\noindent where $\tau$ is the joint torque, k$_{p}$ and k$_{d}$ are the spring and damper coefficients, $\theta$ and $\theta _d$ are the current and desired joint angles, $\dot \theta$ and $\dot \theta _d$ are the current and desired angular velocities.  The torque produced by the PD controller is linearly proportional to the displacement error.  The elastic-damper gains for each DOF joint of the articulated skeleton were determined through human intervention to achieve the necessary responsiveness (see Table ~\ref{table:kp_ks_table}).  Another approach, that we \textit{do not} use, could be an off-line optimization search approaches to calculate the elastic-damper gain coefficients, as done by Geijtenbeek et al. ~\cite{GPS12}.  Additionally, the associated PD coefficients for each joint were scaled by the inertia-rotation (i.e., an inertia-scaled PD-servo) in an attempt to produce less stiff-looking motions ~\cite{LYPS*10,ZMMS+07}.  Note that in addition to the character's ground contact constraint (i.e., the feet) the articulated rigid body's motion is driven solely by joint torques.
\begin{table}
  \centering
  \includegraphics[width=0.8\columnwidth]{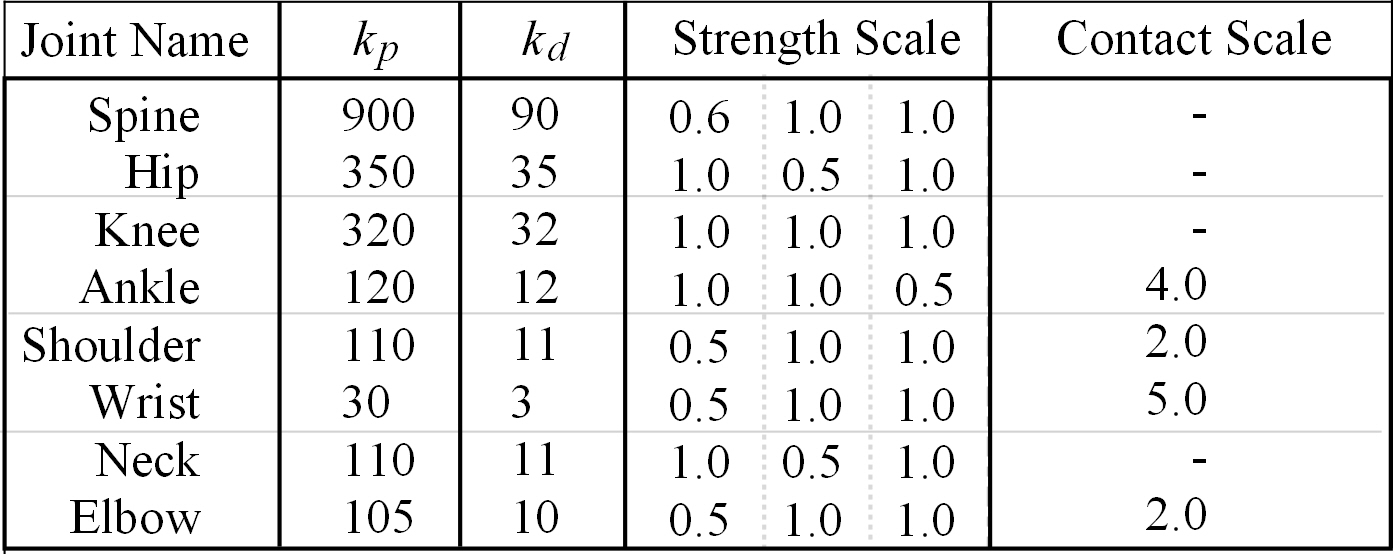}
  \caption{The character model's default joint PD coefficients.}
  \label{table:kp_ks_table}
\end{table}

\subsection{Comfort Factors}
While the basic IP model with feet can produce motions that respond to disturbances, the final pose after a corrective step can look uncomfortable and unnatural.  Hence, we include a secondary condition which identifies if the character is in a comfortable pose (i.e., feet are cross-legged and facing the desired direction).  We accomplish this by adding a desired pelvis direction from which we position the left and right foot locations during comfort stepping (i.e., to the left and right of the character's facing direction).

The comfort decisions are based on the final balanced character pose.  For example, whereby after the character takes a corrective step to recover from a disturbance it could be left standing in an uncomfortable and unnatural pose (e.g., twisted pose or having crossed legs).  The comfort factor did not encompass tweening poses (i.e., the in-between interpolated movements) during stepping transitions (e.g., trajectories could take routes that might not be desirable, such as passing through the support leg).


\figuremacroW
{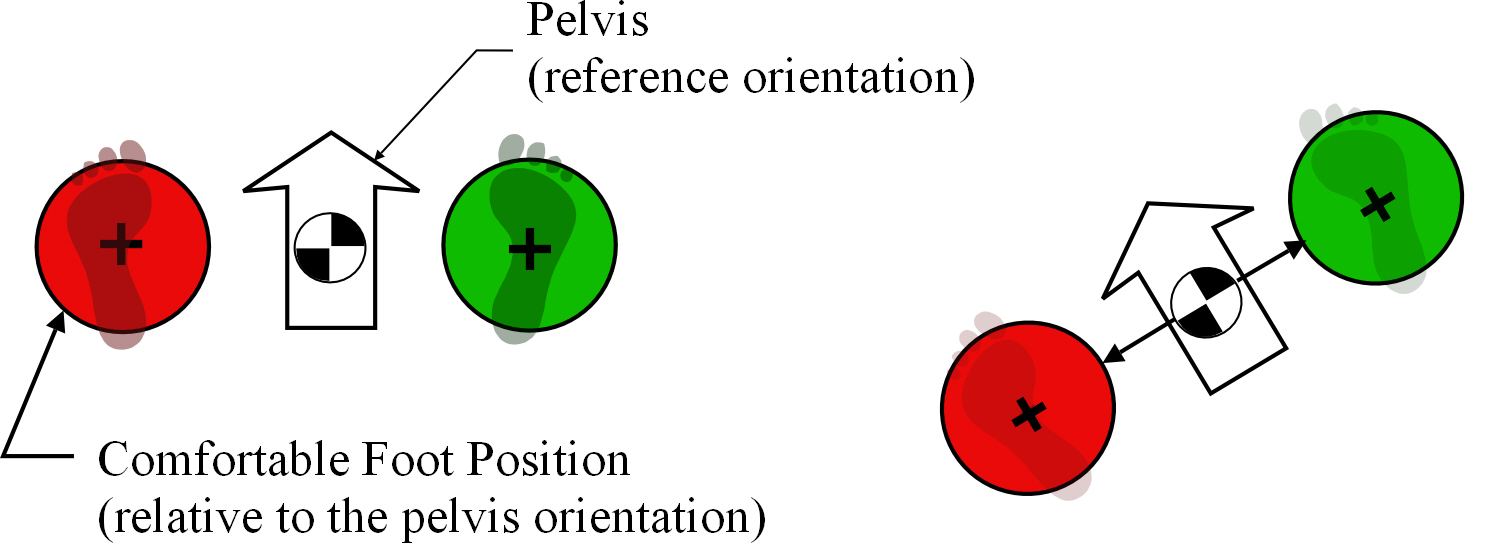}
{Direction}
{We impose a desired direction to the character from which we calculate comfortable foot placement positions and orientations (note, base direction provides a steering/turning control reference mechanism).}
{0.78}

\subsection{Inverse Kinematics}
The inverse kinematic (IK) solver maps a solution between our low-dimensional model and our highly articulated biped skeleton hierarchy.  While the highly articulated skeleton contains a huge amount of flexibility and ambiguity (i.e., multiple solutions for achieving the same goal), in comparison to the simplified low-dimensional model which is minimalistic, computationally efficient, and straightforward to solve.  The simplified model, however, possesses multiple attributes (i.e., overall center-of-mass position and feet locations) that are common to the highly articulated skeleton, which are fundamental for generating physically correct balanced biped stepping poses.  To accomplish the mapping efficiently, we subdivided the IK problem into two separate parts (i.e., upper and lower body).  This made solving the IK problem faster and more robust.  Moreover, our adaptive stepping technique solves balancing logic while the upper-body motions are left free for alternative actions, such as personality and style (e.g., looking around, arms' swaying). 

We focused on lower body movements since they are the most crucial for upright balancing motions ~\cite{TLCL+10} compared to the upper body.  While, foot trajectories were generated by interpolated Bezier splines between the current and desired landing positions during foot transitions.

The final motions \textbf{did not} use any motion capture or key-framed libraries.  Hence, some of the motions may have appeared to look a bit robotic.  This approach can be remedied by combining the generated motions with a multiple priority IK solution (i.e., with a primary and secondary goal).  Whereby, the primary balanced physically correction motion are always enforced, while the secondary less crucial aesthetically pleasing life-like motions are combined on top from sources, such as key-framed libraries or random motion generators.

\section{Experimental Results}
We applied our approach to different simulation situations to demonstrate the advantages of our method and its potential for creating responsive motions without pre-recorded animation libraries (i.e., key-framed data).  The simulations were ran at 100 frames per second (fps) and were performed on an Intel Core i7-2600 CPU with 16-GB of memory running Windows-7 64-bit on a desktop PC.  The results are shown through a series of experiments to warrant the practicality and robustness of our approach for generating adaptive biped stepping motions without key-framed data.  In short, the visual results testify to the highly dynamic and responsive nature of our approach for synthesizing reactive balancing stepping actions.  The overall computational time for generating the character motions with control including dynamic simulation overheads (e.g., rigid body constraints and contacts) was on average less than 5 ms, respectively.

\subsection{Push Disturbances}
While the character was standing upright (e.g., standing or walking) we applied various push disturbances (i.e., random impulse forces to the upper body) from different horizontal directions at random times.  Whereby, for small pushes the character would sway and compensate through ankle torque adjustments, while for larger pushes the character would need to take a corrective step to compensate and remain upright and balanced (e.g., see Figure \ref{fig:push_type}).  The random force disturbances lasted between 0.1s to 0.5s and had a magnitude between 10N to 500N.  Additionally, to emulate a more dynamic interactive environment, we simultaneously threw multiple rigid body spheres (with a density of ${\rho}$ = 50kg/m${^3}$) of varying size at the character (shown in Figure \ref{fig:balls}).  The multiple rigid body collisions with the character triggered simultaneous rapid impulse forces that required the biped to take corrective measures to remedy the disruptive situation and remain upright and balanced through stepping and swaying actions (see Figure \ref{fig:walking_pushed}).


\figuremacroW
{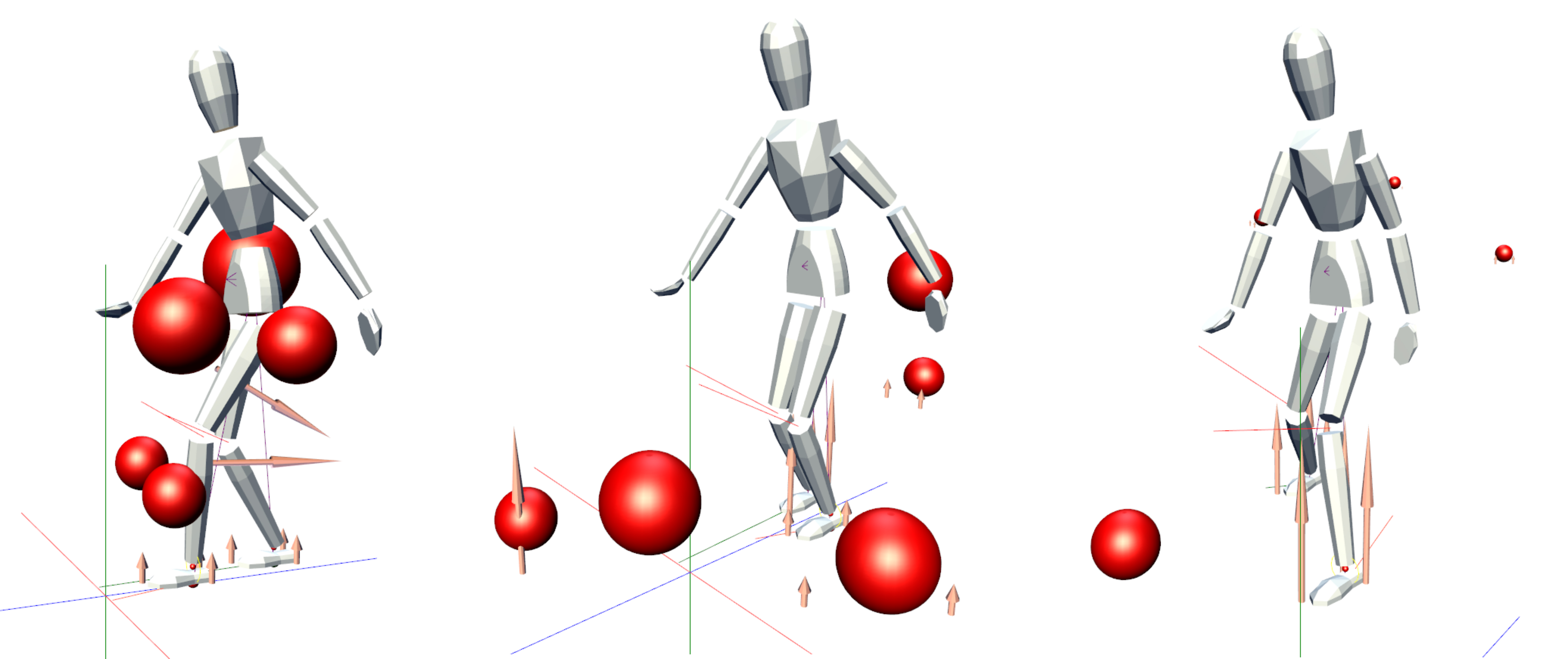}
{Random Balls}
{Multiple rigid body spheres with a density of ${\rho}$ = 50kg/m${^3}$, are fired at the character which causes push forces to be applied to the body and make the character take corrective measures to stay upright and balanced.}
{0.9}


\figuremacroW
{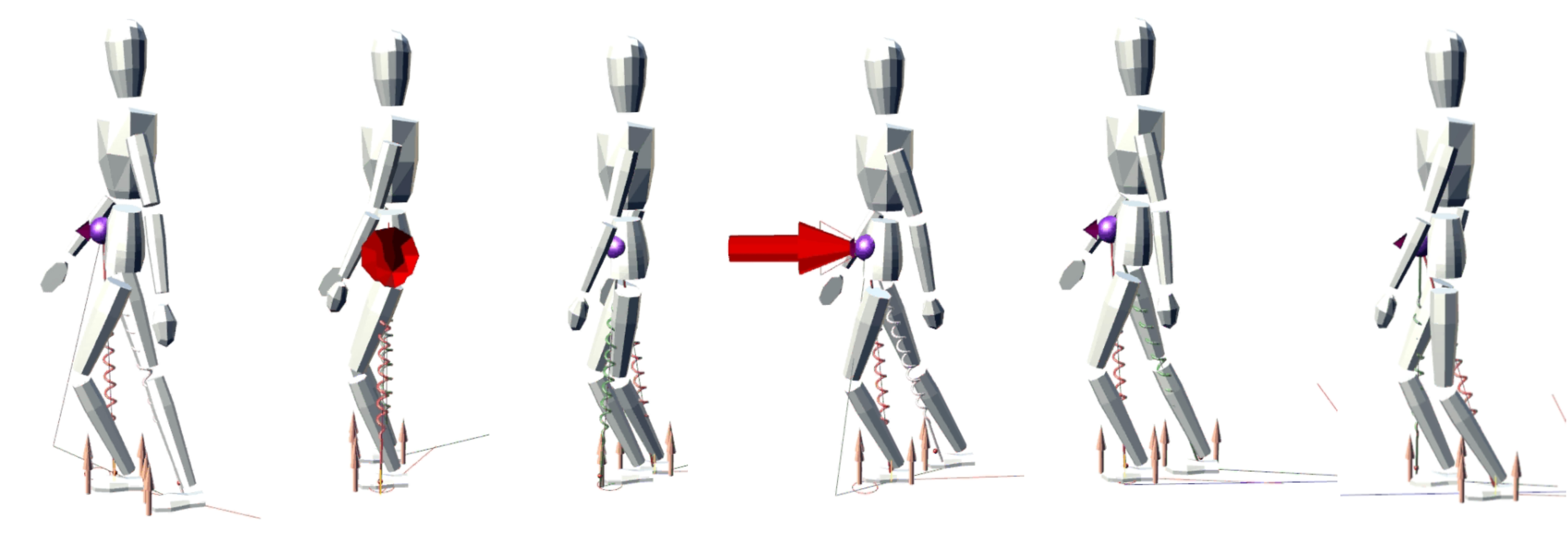}
{Controlled Walking}
{The character walked along a straight line while being repeatedly pushed from different directions at random times (i.e., demonstrating a controlled upright balanced walk while maintaining a constant walk velocity with steering).}
{0.9}

\subsection{Rotating Platform}
\begin{figure}[htbp]
  \centering
  \includegraphics[width=1.0\columnwidth]{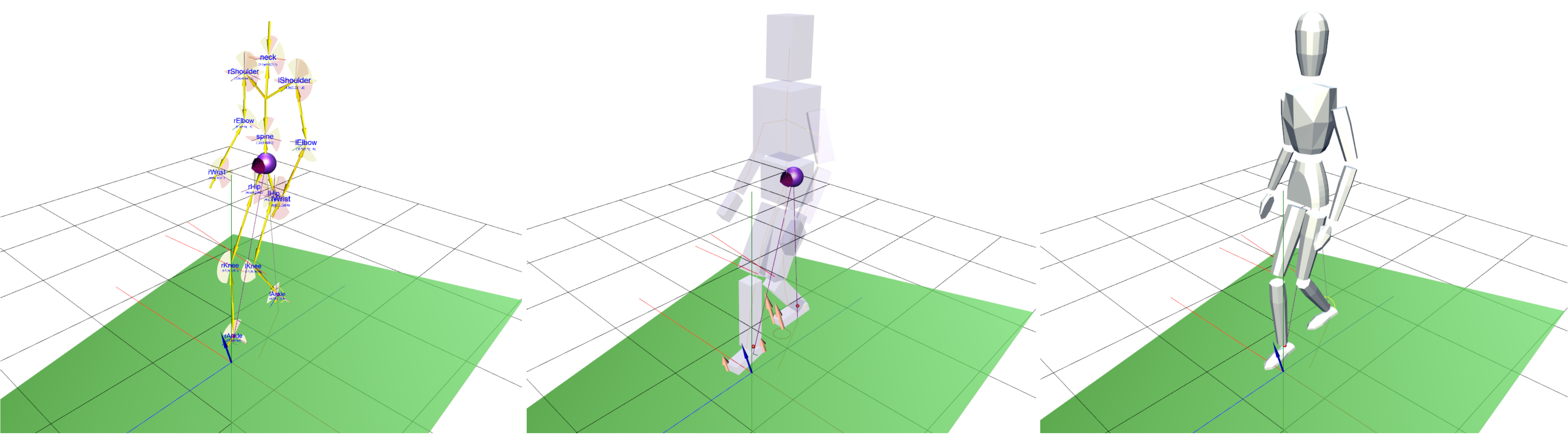}
  \caption{Character remaining balanced while the platform slowly rotates. (Left) Inverse kinematic solution is updated based on the changes in leg lengths.  (Middle) Joint torques are updated for controlling the rigid bodies.  (Right) Graphical solution.}
  \label{fig:legs}
\end{figure}

To show the adaptive nature of our approach for handling changing terrain heights, we placed the character on a continually rotating platform.  The platform was gradually gyrated around the x-z axis in real-time by slopes of up to 45 degrees, while the character constantly adjusted its balance by altering its leg lengths to prevent it falling over.  Furthermore, we extended the test by adding push disturbances to the character while on the rotating platform.  The character would then need to take corrective foot-steps to compensate for the disturbances and shows the robustness of our model (see Figure ~\ref{fig:teaser} and Figure ~\ref{fig:legs}).

\subsection{Holding Boxes}
Since our model is dynamic and the motions are generated on the fly, we had the character hold a box with changing attributes (i.e., mass and size) while staying upright and balanced.  The box's mass was modified in real-time to visually show the character adapting to holding the increased mass through legs bending and posture adjustments while push disturbances were applied (see Figure ~\ref{fig:holding_box} and Figure ~\ref{fig:holding_box_and_pushed}).


\figuremacroW
{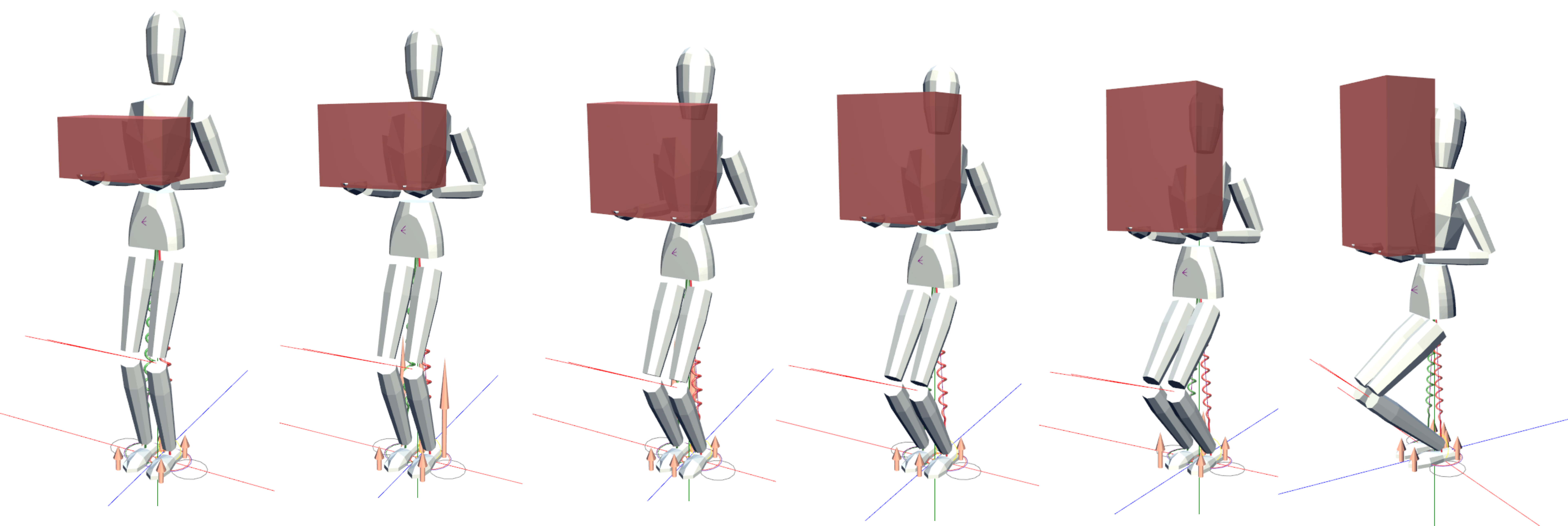}
{Holding A Box}
{The character holding a box while the box's weight is gradually increased in real-time causing the character to adjust its pose and visually show stress (note, the legs' spring stiffness strength is kept constant).  The problem of knee bending can be an unnatural response to carrying a heavy box, where it might be desirable to keep the knees as straight as possible which therefore requires less joint torque.  This can be remedy this by scaling the spring-damper constants based on the overall mass of the character (i.e., adding the weight of the box).  This also accommodates dynamically changing the character's mass at run-time}
{1.0}


\figuremacroF
{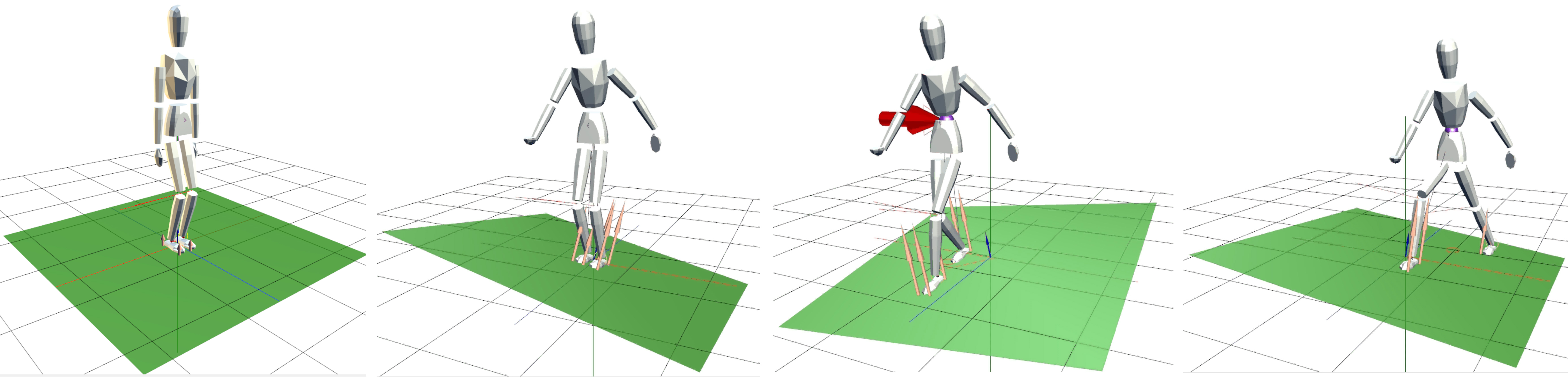}
{Rotating Plane}
{Our adaptive biped stepping approach ensures our character remains upright and balanced during a wide range of interactive situations.  In this example, the ground is constantly rotating while push disturbances are applied, which cause the character to take corrective steps and shift force between the support feet to avoid falling over.}
{1.0}



\subsection{Character Dimensions}
An advantage of our model is its ability to adapt to changes in a character's morphology (e.g., leg lengths) while still being able generate visually plausible, physically correct upright motions.  For example, while keeping the upper-body the same we shortened and lengthened the leg lengths as shown in Figure ~\ref{fig:walking_leg_sizes}.  We found that our control framework was able to adapt the motions to fit in with character dimension changes while the new motions still appeared characteristically correct.

\section{Limitations}
Our model focuses on upright balanced motions, such as standing and walking.  The stepping model does not include friction for the feet (i.e., no slipping).  We use less accurate but more computationally faster approximations to achieve visually pleasing interactive results (e.g., circular foot support areas, bounding boxes for collisions, massless legs).  Additionally, to achieve the desired leg stiffness a user must manually tune the spring-damper coefficients.  Finally, our model only generates key motions based on balancing logic and does not encapsulate any behavioral attributes, such as arm movements (i.e., raising their arms to protect themselves), emotions (i.e., happy, tired, sad).  However, research has been done to couple physical reactions with animation libraries in an attempt to synthesize the behavioural aspects of a characters response which could be combined with our method ~\cite{ZMCF05,KHL05,AFB05}.  Our approach uses the hip midpoint as the COM position similar to SIMBICON ~\cite{YLP07} due to it being fast and simple, however, the model could be adapted to constantly update and track the full articulated body COM position synonymous with the approach by Tsai et al. ~\cite{TLCL+10}.

While our results appeared human-like and responsive, the results from our generated locomotive animations were not as visually pleasing as those created by motion capture libraries (e.g., just walking along a straight-line).  Our controller appeared to produce motions with stiffer balancing corrections.  A possible explanation for this was that our model did not include corrective forces from the heel and toe and did not use any upper body postural swaying for balancing, instead only relying on the lower-body.  One possible solution would be to include further adaptations to the model; or alternatively, combine the generated base motions with locomotive data (e.g. pre-recorded animations) as done by Yin et al. ~\cite{YLP07} and Coros et al. ~\cite{CBM10}.

Our proposed approach only simulates the lower body movements, which can causes the simulated character to not always react to environmental variations naturally like a human.  For example, the demonstrated results in Figure \ref{fig:holding_box} may not appear to look realistic; for example, when lifting a heavy box, we do not bend our keens too much, while the pose in Figure \ref{fig:holding_box_and_pushed} may not visually display the strenuous nature of the push and the weight of the box.

The joint torques are on occasion quite high (i.e., ~500Nm, similar to the values given by Lee et al. \cite{LKL10}) and so could on occasion produce stiff motions; however, one approach would be to incorporate a feed-forward torque compensation for gravity as done by Mordatch et al. \cite{MLH10} to reduce the PD gains.


\figuremacroF
{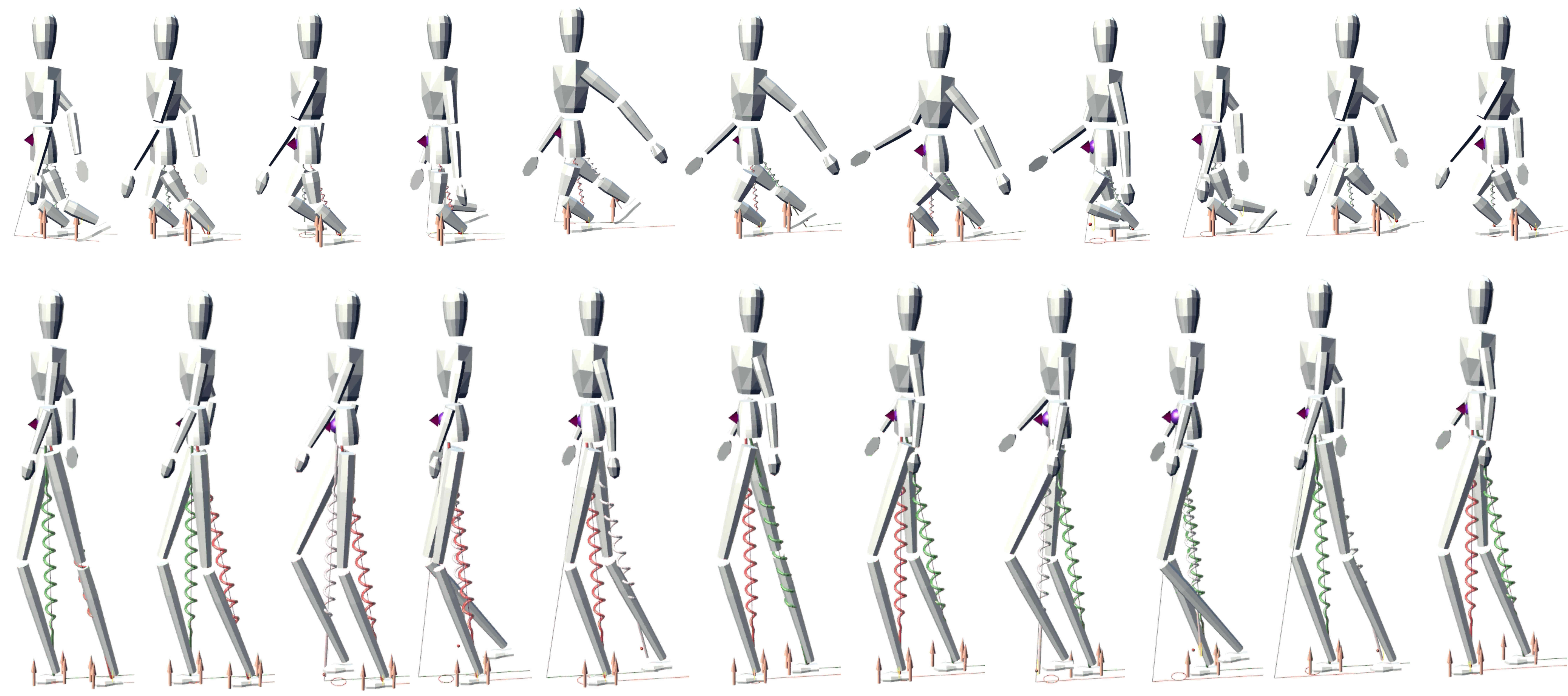}
{Leg Lengths}
{The model automatically adapts to different leg sizes.}
{1.0}

\section{Discussion}
The paper describes the evolution of the spring loaded inverse pendulum control into a cooperative set of techniques using PD and inverse kinematics to a 36-DOF character for dynamic control of upright motions (e.g., \textit{standing} and locomotion).  The method runs in real-time, is relatively easy to implement, and is robust to different dynamic configurations (such as terrain height deviations, pushes, and allowing the character to hold objects of varying masses).  The paper extends the oversimplified IP model (i.e., constantly stepping) to include \textbf{ankle-torque feedback} for control and balanced \textbf{standing-without stepping}, swaying for small disturbances without needing to take steps, steering, and comfort factors (e.g., avoiding cross-legged and feet overlapping).  The character motions are \textbf{self-driven} and \textit{do not} use any data-driven sources (e.g., motion capture data).

The work exploited the SLIP-base IP approach to emphasis a less rigid, flexible, and dynamic approach for generating upright motion (i.e., naturalness/robustness/speed).  Since the SLIP with damping incorporates swaying, allow leg length deviations, computationally fast, and sufficiently stable and robust (i.e., if adequately dampened).

\figuremacroW
{holding_box_and_pushed}
{Holding A Box While Being Pushed}
{Our adaptive biped stepping approach ensures our character remains upright and balanced during a wide range of interactive situations.  For example, the figure shows our character holding a box, while we repeatedly apply random push forces; requiring the character to take a corrective step to regain its balance and not fall over.}
{1.0}

\section{Conclusion}
We have presented a computationally fast and straightforward technique for synthesizing responsive upright 3D biped character stepping motions that can handle a range of unanticipated disturbances, such as dynamic terrain and push perturbations.  While our technique focused on lower body motions, it could, nevertheless, handle both small and large push forces even during terrain height variations.  Moreover, our model effectively created human-like motions that synthesized low-level upright balancing movements.  We anticipate that our approach could be combined with additional controller techniques to produce whole body autonomous agents.  For example, one promising direction would be the combination of scripted coherent random movements into the final system so the final motions are responsive and life-like and possess non-repetitive personality qualities (e.g., worried, nervous, tired, bored) described by Perlin ~\cite{Ken95}.

This paper has only scratched the surface of the subject of generating intelligent physics-based character motions and is an active area of research.  While our model would still need to include a diverse range of actions to make it a viable solution for online applications (e.g., sit, kick, get-up), we believe, however, that our model is flexible enough to be combined with hybrid techniques or additional physics-based controllers to produce human-like avatars that can unconsciously handle unpredictable situations that mimic the real-world.

Alternatively, it should be mentioned that our approach could potentially be applied to other animation areas, for example motion capture editing.  Whereby, an animator could use actual motion capture sequences and compare them in real-time with our models predicted results to help adapt and identify physically inaccurate animations.  Since our model would be useful in automatically identifying realistic foot-ground contact positions during dynamic interactions.  Hence, the final motions would contain the detailed personality of the motion capture data while permitting the animator to correct stepping and balancing details.

There is an increasing inclination towards more innovative and novel approaches of generating virtual character animations that go beyond basic data-driven techniques using unintelligent puppet-like rag-doll physics models towards more logic-driven smart solutions.  Whereby, these smarter solutions attempt to emulate the real world and produce more engaging life-like motions.  Hence, moving towards these innovative solutions, like the one we present in this paper, will result in a new generation of immersive and interactive environments with realistic and responsive characters in them.


\figuremacroF
{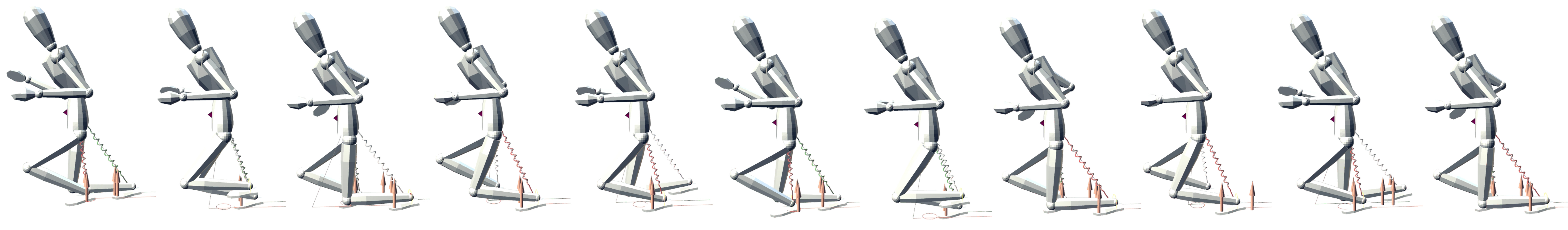}
{Crouching}
{Adjusting the leg parameters can create various stylistic walking types; for example, the figure shows the character crouched while walking due to the default leg length being halved.}
{1.0}

\figuremacroW
{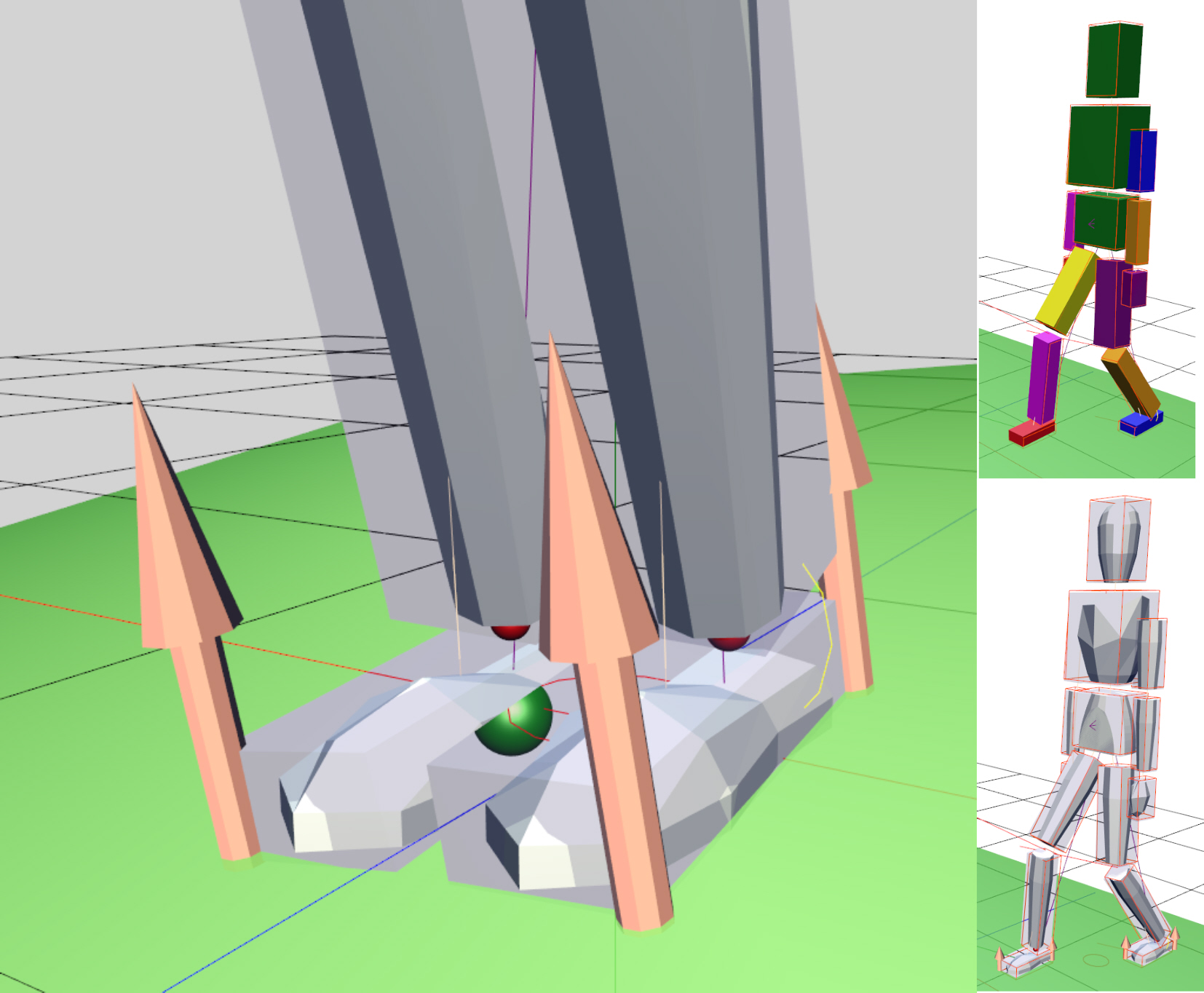}
{Orientated Bounding Boxes}
{The articulated rigid-body model was decomposed into low-dimensional orientated bounding boxes for contact information with the environment.}
{1.0}

\section*{Acknowledgements}
We are indebted to the reviewers for taking the time out of their busy schedules and for providing invaluable comments and suggestions to help to improve the quality of this paper.



\small

\bibliographystyle{IEEEtran}
\bibliography{paper}



\end{document}